\documentclass{article}
\pdfoutput=1
\usepackage{jcappub}
\usepackage[normalem]{ulem}
\usepackage{amsmath}
\usepackage{amssymb}
\usepackage{xcolor}
\usepackage{hyperref}
\usepackage{slashed}
\usepackage{braket}
\usepackage{physics}
\usepackage{comment}
\usepackage[normalem]{ulem}
\usepackage{slashed}

\newcommand\underrel[2]{\mathrel{\mathop{#2}\limits_{#1}}}

\newcommand{\edit}[1]{\textcolor{black}{#1}}

\begin{document}

\title{Warm inflation with a heavy QCD axion}

\author[a,b]{Kim~V.~Berghaus,}
\emailAdd{berghaus@caltech.edu}

\author[b,c]{Matthew Forslund,}
\emailAdd{matthew.forslund@stonybrook.edu}

\author[b]{Mark Vincent Guevarra}
\emailAdd{markvincent.guevarra@stonybrook.edu}

\affiliation[a]{Walter Burke Institute for Theoretical Physics, California Institute of Technology, 1200 E California Blvd, Pasadena, CA 91125, USA}
\affiliation[b]{C. N. Yang Institute for Theoretical Physics, Stony Brook University, 100 Nicolls Rd, Stony Brook, NY 11794, USA}
\affiliation[c]{Department of Physics, Brookhaven National Laboratory, 98 Rochester St, Upton, New York 11973 USA}

\abstract{We propose the first model of warm inflation in which the particle production emerges directly from coupling the inflaton
to Standard Model particles. 
Warm inflation, an early epoch of sustained accelerated expansion at finite temperature, is a compelling alternative to cold inflation, with distinct predictions for inflationary observables such as the amplitude of fluctuations, the spectral tilt, the tensor-to-scalar ratio, and non-gaussianities. 
In our model a heavy QCD axion acts as the warm inflaton whose coupling to Standard Model gluons sources the thermal bath during warm inflation.  
Axion-like couplings to non-Abelian gauge bosons have been considered before as a successful microphysical theory with emerging thermal friction that can maintain finite temperature during inflation via sphaleron heating. However, the presence of light fermions charged under the non-Abelian group suppresses particle production, hindering a realization of warm inflation by coupling to QCD. 
We point out that the Standard Model quarks can be heavy during warm inflation if the Higgs field resides in a high-energy second minimum which restores efficient sphaleron heating. A subsequent large reheating temperature is required to allow the Higgs field to relax to its electroweak minimum. 
 Exploring a scenario in which hybrid warm inflation provides the large reheating temperature, we show that future collider and beam dump experiments have discovery potential for a heavy QCD axion taking the role of the warm inflaton. 

  }

\maketitle

\section{Introduction }
\label{sec:intro}
Cosmic inflation, an early period in which the universe undergoes accelerated expansion, is a successful explanation of the observed near homogeneity and isotropy in the cosmic microwave background (CMB) \cite{PhysRevD.23.347, LINDE1982389, PhysRevLett.48.1220}. 
Yet, the description of the energy content that induces cosmic inflation and its transition to a radiation dominated universe filled with a quark-gluon plasma remains an active topic of investigation \cite{PhysRevLett.73.3195, PhysRevD.42.2491, PhysRevD.51.5438, ABBOTT198229, PhysRevLett.76.1011,Takahashi:2019qmh,  Figueroa:2020rrl,Takahashi:2021tff, Figueroa:2021yhd,Narita:2023naj}. 

Many models assume a slowly rolling scalar field, the inflaton, at zero temperature that in a subsequent stage reheats the universe to finite temperature \cite{Workman:2022ynf}. 
Quantum fluctuations of the inflaton seed the large scale structure observed in the CMB anisotropies and the slow-roll softly breaks scale-invariance, explaining the observed spectral tilt in the CMB power spectrum. 
However, the simplest models are ruled out by measurements constraining the tensor-to-scalar ratio $r$ \cite{Planck:2018jri}. 

Warm inflation is a compelling alternative, in which cosmic inflation occurs at finite temperature \cite{Berera:1995ie, Berera:1995wh, Berera:1999ws, Berera:1998px, Berera:2008ar, Bastero-Gil:2016qru, Kamali:2023lzq, Berera:2023liv}. 
If the inflaton efficiently sources a thermal bath of particles through thermal friction such that the production rate compensates for the dilution of Hubble expansion, a quasi steady-state temperature can be maintained. The finite temperature enhances fluctuations that seed large scale structure, which modifies the prediction for CMB power spectra and suppresses contributions to the tensor-to-scalar ratio $r$. 
Due to the thermal friction 
the field range the inflaton needs to transverse to support 60 e-folds of inflation can be sub-Planckian.
Additionally, a  separate reheating stage may not be required; inflation can end when the radiation starts dominating over the energy density of the inflaton.
Interestingly, warm inflation predicts enhanced non-gaussianities with a distinct bispectral shape \cite{Bastero-Gil:2014raa, Mirbabayi:2022cbt}, a 'smoking gun' that may become accessible within the next decade of experiments, making this an opportune time to investigate compelling models of warm inflation.  

A key ingredient for a successful warm inflation model is a microphysical theory in which efficient particle production emerges from the fundamental particle interaction between the inflaton field and the light particles it couples to \cite{Berera:1998px, Berera:2008ar, Berera:2023liv}. 
Historically, this was a challenge due to dominant thermal mass corrections accompanying the emerging thermal friction \cite{Yokoyama:1998ju}, which necessitated theories with many fields \cite{Berera:1999ws}; however since warm inflation was first proposed thirty years ago, much progress has been made e.g.~\cite{Berera:2008ar, Bastero-Gil:2016qru, Berghaus:2019whh}.
In particular, in 'minimal warm inflation' \cite{Berghaus:2019whh}, recently referred to as a flawless microscopic realization of the idea \cite{Mirbabayi:2022cbt}, the inflaton is an axion-like particle (ALP) coupling to non-abelian gauge fields which induces particle production via sphaleron heating.
The thermal back-reaction is suppressed due to the (approximate) shift symmetry of the ALP, and efficient particle production is ensured through sphaleron heating, which we review in detail in Sec.~\ref{sec:Microphysics}.

In this work, we propose a heavy QCD axion as the inflaton of warm inflation with QCD gluons in their non-confined phase constituting the radiation component.
Sphaleron heating is suppressed in the presence of light quarks \cite{Berghaus:2020ekh}, thus in order for the QCD axion to be viable for warm inflation the light Standard Model (SM) quarks need to be heavy during the inflationary phase.
We construct a model in which quark masses are much larger during inflation and subsequently relax to their measured values after inflation concludes. 
Our work describes the first successful implementation of warm inflation in which the thermal friction emerges directly from coupling the inflaton, here a heavy QCD axion, to SM particles. 

The structure of this paper is as follow. In Sec.~\ref{sec:overview} we provide an overview over the dynamics of warm inflation, highlighting differences from cold inflation and their impact on the theoretical description and observables. In Sec.~\ref{sec:Microphysics} we discuss how coupling the inflaton to other particles gives rise to particle production as well as thermal masses, and summarize the proposed microphysical models in the literature. We go over the particle production mechanism due to sphaleron heating in minimal warm inflation in Subsec~\ref{subsec:MWI}, and pedagogically review how the presence of light quarks suppresses sphaleron heating, posing an obstacle towards using the coupling to QCD as the particle production mechanism in Subsec.~\ref{subsec:MWIwQCD}. 

In Sec.~\ref{sec:MWIQCD} we introduce the UV ingredients of our minimal warm inflation model with a heavy QCD axion that allow quark masses to be heavy in the early universe to restore sphaleron heating, and a heavy QCD axion that can act as the inflaton. In Subsec.~\ref{subsec:SMvacua} we discuss how additional UV Higgs vacua that raise quark masses emerge when a dimension six operator stabilizes the negatively running quartic coupling of the Higgs field at high energies. 
In Sec.~\ref{sec:heavyQCDaxions} we review the motivation for considering extra mass contributions to the QCD axion beyond the usual QCD instanton ones, and discuss models that lead to such contributions while preserving the axion's solution to the strong CP problem. 

Finally, in Sec.~\ref{sec:warmhybridinflation} we consider a concrete warm inflationary scenario that is compatible with all inflationary observables, and features a large reheating temperature that  
allows the Higgs field to relax to its electroweak (EW)-vacuum once warm inflation has concluded. Mapping out the predicted parameter space for the warm heavy QCD axion inflaton in the $f - m_{\phi}$ plane, we discuss current constraints and future discovery potential in Sec.~\ref{sec:constraints}. Lastly, we conclude with a summary of our findings and outline future directions in Sec.~\ref{sec:concl}.

While this work was in its final stages \cite{Drewes:2023khq} appeared which analyzed how scattering processes in an interacting plasma can provide chirality violations that restore sphaleron heating. We leave an application of their result to the SM for future work, and assume that large quark masses are necessary for efficient QCD sphaleron heating.

\section{Overview on Warm Inflation}
\label{sec:overview}
We begin with a brief overview of warm inflation, highlighting how it differs from the standard cold inflationary scenario.    
In the usual paradigm, inflation is driven by a scalar inflaton field with a Lagrangian
\begin{equation}
    \mathcal{L} = \frac{1}{2} g^{\mu \nu}\partial_\mu \phi \partial_\nu \phi - V(\phi) \, ,
\end{equation}
that has a potential $V(\phi)$ driving accelerated expansion in the universe. Decomposing the scalar field into a smooth background component and spatially varying perturbations $\phi(t,\vec{x}) = \phi(t) +\delta \phi(t,\vec{x})$, the 
 homogeneous field $\phi(t)$ in a flat FLRW background
obeys the equation of motion (EOM)
\begin{equation}
    \ddot{\phi} + 3H \dot{\phi} + V'(\phi) = 0 \, .
\end{equation}
The Friedmann equation is determined by the energy content of the scalar field
\begin{equation}
    H^2 = \frac{1}{3M_{\text{Pl}}^2} \left( \frac{1}{2}\dot{\phi}^2 + V(\phi) \right) \, ,
\end{equation}
where $H = \frac{\dot{a}}{a}$, is the Hubble parameter and $a$ is the scale factor. 
Any energy density of additional particles besides the scalar field is usually assumed to be negligible, as even if at some initial time there was a sizeable component, Hubble expansion would dilute it in an exponentially expanding universe.  
Consequently neglecting other particles during inflation is often an excellent approximation, and standard cold inflation occurs at near zero temperature such that any spatial fluctuations around the homogeneous background $\delta \phi(t,\vec{x})$ are quantum fluctuations.  

In order to deviate from this cold inflationary picture, the scalar field has to continuously source other particles at a rate that at least can keep up with the rate of dilution from Hubble expansion. In particular, if the produced particles also have efficient self-interactions they can constitute a thermal bath with a well defined temperature $T > H$. In a thermal environment with a slowly evolving scalar field compared to the interaction rate of the particles in the thermal bath, the effects of particle production on the classical EOM of the scalar field are to leading order captured by the addition of a linear friction term with a thermal friction coefficient $\Upsilon$ \cite{Laine:2016hma}. Such a macroscopic linear friction coefficient can emerge from a fundamental interaction between the scalar field and particles constituting the thermal bath as we will see below. 

Keeping the interaction general for now, we denote the additions to the Lagrangian as \cite{Laine:2016hma} 
\begin{equation}
    \mathcal{L} = \frac{1}{2}g^{\mu \nu }\partial_\mu \phi \partial_\nu \phi - V(\phi) - \phi \mathcal{J}_{\text{int}} + \mathcal{L}_{\text{bath}} \, ,
\end{equation}
such that the homogenous EOM of the scalar field has an additional term,
\begin{equation}
    \ddot{\phi} + 3H \dot{\phi} + V'(\phi) = -\langle \mathcal{J}_{\text{int}} \rangle^{\text{th}}_{\text{non-eq}}  \, ,    
\end{equation}
and $\mathcal{L}_\text{bath}$ allows for efficient self-interactions of the particles in the bath, ensuring thermalization. 
Here $\langle \mathcal{J}_\text{int} \rangle^{\text{th}}_{\text{non-eq}}$ denotes the thermal expectation value of the operator $\mathcal{J}_{\text{int}}$ in the presence of the coupling to the scalar field which gives it a nonzero value, compared to when the system is in thermal equilibrium $\langle \mathcal{J}_{\text{int}} \rangle_{\text{eq}}^{\text{th}} = 0$. Linear response theory  relates $\langle \mathcal{J}_{\text{int}} \rangle_{\text{non-eq}}^{\text{th}}$ at leading order to the properties of the scalar field $\phi$, $\dot{\phi}$, and to macroscopic quantities such as the thermal mass $m^2_{\text{th}}$ and the thermal friction coefficient $\Upsilon$,  
\begin{equation} \label{eq:linear_response}
\langle \mathcal{J}_{\text{int}}\rangle_{\text{non-eq}}^{\text{th}} = m^2_{\text{th}} \phi + \Upsilon \dot{\phi} + \mathcal{O}(\dot{\phi}^2,\nabla^2 \dot{\phi},(\nabla \phi)^2, \ddot{\phi}) \, ,
\end{equation}
which are determined by thermal correlators of the equilibrium theory \cite{Laine:2016hma}. Thus, once $\mathcal{J}_{\text{int}}$ is specified one can calculate the thermal friction coefficient $\Upsilon$ from first principles, and the classical EOM of the scalar field simplifies to 
\begin{equation}
    \ddot{\phi} + (3H + \Upsilon) \dot{\phi} + V'_{\text{eff}}(\phi) = 0  \, ,   
    \label{eq:EulerLagrangeWarmInflation}
\end{equation}
where $V'_{\text{eff}} = V'({\phi}) + m^2_{\text{th}} \phi$. Here the energy density extracted from the scalar field via $\Upsilon$ goes into sourcing the thermal bath of particles such that continuity equation for the radiation component is given by
\begin{equation}
\label{eq:EOMradiation}
   \dot{\rho}_R + 4H \rho_R = \Upsilon \dot{\phi}^2 \, . 
\end{equation}
The temperature of the thermal bath is determined by the number of thermalized degrees of freedom $g_*$ 
\begin{align}
    \rho_R = \frac{\pi^2}{30} g_* T^4 \, . \label{RadiationEnergyDensityWI}
\end{align}
In this warm inflationary picture the Friedmann equation has an additional term that accounts for the radiation energy density,  
\begin{align}
    H^2 = \frac{1}{3M_{\text{Pl}}^2} \left( \frac{1}{2}\dot{\phi}^2 + V_{\text{eff}}(\phi) + \rho_R \right) \, . \label{HubbleConstantWarmInflation} 
\end{align}
Satisfying the following slow-roll conditions guarantees an extended period of warm inflation, 
\begin{align}
    \varepsilon_H \equiv -\frac{\dot{H}}{H^2} \ll 1 \, , \label{EpsilonSlowRoll} \\
    \eta_H \equiv -\frac{\ddot{H}}{\dot{H}H} + \frac{\ddot{\phi}}{H\dot{\phi}} \ll 1\, \label{EtaSlowRoll} \, . 
\end{align}
Additionally, the thermal mass correction needs to be small (e.g. $V_{\text{eff}}(\phi) \approx V(\phi)$, $m^2_{\text{th}} \phi \ll \Upsilon \dot{\phi}$) to avoid a large temperature dependent back-reaction on the inflaton potential. We will discuss this requirement 
in detail in Subsec.~\ref{sec:Microphysics}. 
Then, using Eq.~\eqref{HubbleConstantWarmInflation} and imposing the slow-roll conditions \eqref{EpsilonSlowRoll}, \eqref{EtaSlowRoll}, 
 we can neglect $\ddot{\phi}$ in Eq.~\eqref{eq:EulerLagrangeWarmInflation}, and neglect $\frac{1}{2}\dot{\phi}^2, \,\text{and}\, \dot{\rho_R}$ in Eq.~\eqref{eq:EOMradiation}, simplifying the Friedman equation, the EOMs of the scalar field, and the continuity equation of the thermal radiation to  
\begin{align}
     H^2 \approx & \, \frac{1}{3M_{\text{Pl}}^2} V(\phi) \, , \label{eq:SlowRollApprox2} \\
    {\dot{\phi}} \approx & -\frac{V'(\phi)}{3H (1+Q)} \, , \label{eq:SlowRollApprox1} \\
    4H\rho_R \approx & \, 3HQ \dot{\phi}^2\, ,\label{eq:SlowRollApprox3}
\end{align}
where we have introduced the dimensionless parameter $Q \equiv \frac{\Upsilon}{3H}$. Eq.~\eqref{eq:SlowRollApprox3} indicates that a quasi-steady state is reached with an approximately constant temperature. Differentiating Eqs.~\eqref{eq:SlowRollApprox2} and \eqref{eq:SlowRollApprox1} with respect to time gives slow-roll parameters in terms of the potential,
\begin{align}
    \varepsilon_V  &\equiv \frac{M_{\text{Pl}}^2}{2(1+Q)} \left( \frac{V'(\phi)}{V(\phi)} \right)^2 \simeq \varepsilon_H \, , \label{eq:Vslowroll1} \\
    \eta_V  &\equiv \frac{M_{\text{Pl}}^2}{1+Q} \frac{V''(\phi)}{V(\phi)} \simeq \eta_H + \varepsilon_H \label{eq:Vslowroll2} \, .
\end{align}
As in cold inflation, the potential slow-roll parameter, $\epsilon_V$, relates the amounts of e-folds $N$ to a field range in $\Delta\phi$, with the additional factor of $\sqrt{1+Q}$, accounting for the thermal friction:  
\begin{align}
    N(\phi_{*}) \equiv \int_t^{t_{\text{end}}} H dt  &=  \frac{1}{M_{\text{Pl}}^2} \int_{\phi_{\text{end}}}^{\phi_{*}} \frac{V(\phi)}{V'(\phi)} (1+Q) d\phi \, ,\\    & =\frac{1}{M_{\text{Pl}}}\int_{\phi_{\text{end}}}^{\phi_*} \sqrt{\frac{1+Q(\phi)}{2\epsilon_V(\phi)}} d\phi \, .
    \label{eq:NFoldsWarmInflation}
\end{align}
In the weak regime of warm inflation $Q \ll 1$, the corrections to the background evolution of the scalar field are small. However, the presence of the thermal bath of particles still changes the nature of the primordial density perturbations to be of classical origin, rather than seeded by quantum fluctuations. The quantum inflaton perturbations in k-space scale as $\delta \phi_k \sim H$ in cold inflation, whereas the classical inflaton perturbations in warm inflation scale as $\delta \phi_k \sim \sqrt{H T}$.   
Thus, the finite temperature $T$ enters as a new scale that enhances perturbations.

In the strong regime of warm inflation $Q \gg 1$ 
a large amount of e-folds is achieved over a smaller field range. The thermal inflaton perturbations in the strong regime are even more efficiently  enhanced compared to cold inflation by the temperature $T$, as well as $Q$. The parametric dependence of the scaling in which $Q$ enters depends on the temperature dependence of the thermal friction coefficient $\Upsilon(T)$ \cite{2009JCAP...07..013G, Bastero-Gil:2011rva, Das_2020, Mirbabayi:2022cbt}. Due to this strong enhancement of the scalar power spectrum, the tensor-to-scalar ratio, $r$, is very suppressed in the strong regime of warm inflation.

\edit{The predicted spectral tilt in warm inflation also differs from the standard cold inflation result, 
\begin{equation}
n_s -1 = 2 \eta_V -6 \epsilon_V \ \, .
\end{equation}
In the weak regime corrections can remain subleading of order $\mathcal{O}(Q)$ \cite{Ramos:2013nsa}. However, in the strong regime of warm inflation the predicted spectral tilt changes the signs and coefficients in front of $\eta_V$ and $\epsilon_V$. An intuitive explanation is that Hubble decreases during inflation, however the temperature and the friction tend to increase which biases the result towards a blue tilted spectral index. 
The precise result depends on the temperature scaling of the friction coefficient $\Upsilon(T)$, and exhibits weak dependence on $Q$. For example for $\Upsilon \propto T^3$, and $Q > 50$ the spectral tilt simplifies to 
\begin{equation}
n_s - 1 = 8 \epsilon_V - 6\eta_V \, .
\end{equation}
A simple renormalizable power-law potential is not able to fit the observed spectral tilt $n_s \approx 0.965$, as it cannot fullfill $\eta_V \gtrsim 2 \epsilon_V$ \cite{Berghaus:2019whh}. Similar arguments apply in the strong regime also for models which scale as $\Upsilon \propto T$. However, well motivated models such as hybrid inflation for example can easily achieve the necessary hierachy in slow roll parameters without requiring a fine-tuning in the potential \cite{Berghaus:2019whh}.}

Predictions for non-gaussianities are enhanced in warm inflation, and could function as a smoking gun \cite{Bastero-Gil:2014raa, Mirbabayi:2022cbt}. In particular, for a friction coefficient scaling as $\Upsilon \propto T^3$, a distinct new bispectral template characterizes the predicted sizeable non-gaussianity alongside the equilateral bispectral shape \cite{Mirbabayi:2022cbt}.

\section{Microphysics of Warm Inflation}
\label{sec:Microphysics}
A major challenge in constructing a successful microphysical particle model of warm inflation is the large thermal back-reaction on the inflaton potential, a generic feature of models not protected from a thermal mass correction by a symmetry or heavy mediators. For the simple example of an interaction term with massless scalar fields $\chi$ \cite{Yokoyama:1998ju}, $\mathcal{J}_{\text{int}} = \frac{g^2}{2} \phi \chi^2$, the thermal mass correction is $m^2_{\text{th}} = \frac{g^2 T^2}{12} - \frac{ g \phi T}{4\pi }$, the leading term in Eq.~\eqref{eq:linear_response}, in the regime in which $g \phi \ll T$. The friction term $\Upsilon \sim \frac{\phi^2}{ T}$ arises as a correction to the thermal mass term due to its dependence on $\phi$. Thus, while naively it appears that $\Upsilon$ can be large, since its origin is a correction to the thermal mass, it is sub-dominant such that $\Upsilon \dot{\phi} \ll m^2_{\text{th}} \phi$. 
In the scenario in which the potential of the inflaton is such that the thermal mass correction is small, $V \approx V_{\text{eff}}$, the friction term is even smaller, and not be able to play an important role in the dynamics of the scalar field.         In the regime in which the thermal mass correction dominates the inflaton potential $V_{\text{eff}} \gg V$, the total energy density is dominated by radiation impeding the period of accelerated expansion necessary for inflation \cite{Yokoyama:1998ju}. Thus, the required hierachy of  $V_{\text{eff}}(\phi) \approx V(\phi)$, $m^2_{\text{th}} \phi \ll \Upsilon \dot{\phi}$ is not generic.

The first proposal that successfully achieved a small thermal mass correction compared to the thermal friction coefficent,  by Berera in 1999 \cite{Berera:1999ws} suppressed the thermal mass correction by coupling the inflaton to heavy fields, which coupled to the light degrees of freedom. Achieving significant thermal friction required a large number of fields $g_{*} > 10^4$.  In 2016, Bastero-Gil et al.~\cite{Bastero-Gil:2016qru} introduced the 'warm little inflaton', in which the inflaton is a pseudo Golstone boson of a broken gauge symmetry, directly coupling to two light fermions whose thermal mass corrections cancel. 
\subsection{\edit{Sphaleron heating}}
\label{subsec:MWI}
Meanwhile, it was  realized that an axion-like particle is a great candidate for warm inflation \cite{Mishra:2011vh,Visinelli:2011jy,Kamali:2019ppi,Berghaus:2019whh}. In particular, coupling an ALP to non-abelian gauge fields of an SU($N_c$) gauge group via the interaction term
\begin{equation} 
\label{eq:GGdual}
\mathcal{J}_{\text{int}} = \frac{\alpha}{8 \pi f} G^a_{\mu\nu} \widetilde{G}_a^{\mu \nu} \, , \, \,\,\,\,\,\,\,\,\,\, \widetilde{G}^{\mu\nu} \equiv \frac{1}{2} \epsilon^{\mu\nu\alpha\beta} G_{\alpha \beta}\, ,
\end{equation}
can give a large thermal friction coefficient  due to sphaleron heating \cite{Berghaus:2019whh}, while the thermal mass contribution is suppressed by the (approximate) shift symmetry of the ALP.

Sphalerons arise due to the topological nature of the vacuum in non-abelian field theories. Non-abelian theories possess multiple classical vacua which require a large gauge transformation of non-trivial topology to go from one vacuum to another. These different vacua are distinguished by the topological Chern-Simons number $N_{CS}$.
At zero temperature vacuum transitions can occur via instanton processes that allow for tunneling. The non-perturbative mass contribution to an ALP from instantons at zero temperature scales as $m^2_{\phi,\text{inst}}=\frac{\Lambda^4}{f^2}$, where $\Lambda$ is the confinement scale of the non-abelian sector. At temperatures much above the confinement scale, corresponding to weak couplings, $\alpha \ll 1$, instanton processes are negligible, and the non-perturbative mass contribution is highly suppressed by powers of $(\frac{\Lambda}{T})^X$, where lattice simulations have estimated $X \gtrsim 3$ \cite{Frison:2016vuc}. 

However, at  large temperatures, $T \gg \Lambda$, classical transitions between vacua with different topology due to random thermal fluctuations can occur via sphalerons, which are unstable large finite-energy configurations of the gauge fields, which occur on length scales $\sim 1/(\alpha T)$ \cite{PhysRevD.55.6264, KHLEBNIKOV1988885}. The mean-squared fluctuations of the Chern-Simons number are characterized by the sphaleron rate, $\langle N^2_{CS} \rangle \simeq  V t \Gamma_{\text{sph}}$, where $Vt$ is the considered space-time volume \cite{KHLEBNIKOV1988885}.
The sphaleron rate is a well studied quantity due to its relevance in electroweak baryogenesis, and is given by \cite{Moore:2010jd}
\begin{equation}
    \Gamma_{\text{sph}} = \kappa \alpha^5 T^4\, ,
\end{equation}
where $\kappa \sim N^5_c$, up to logs in $\alpha$, and $\mathcal{O}(1)$ factors specified in \cite{Moore:2010jd}. We show the dependence on $\kappa$ and $\alpha$ on temperature due to running in Appendix \ref{app:A} in Fig.~\ref{fig:alphas} for two relevant scenarios.

The rolling ALP inflaton field with $\langle \dot{\phi} \rangle \neq 0$ induces biased real-time sphaleron processes, where $\langle \dot{\phi} \rangle/f$ acts like a chemical potential for the Chern-Simons number $N_{CS}$ \cite{Berghaus:2020ekh}. The ongoing biased vacuum transitions via sphalerons lead to continuous particle production of gauge bosons. 
Plugging Eq.~\eqref{eq:GGdual} into Eq.~\eqref{eq:linear_response}, we find \cite{Berghaus:2020ekh}
\begin{equation}
\Big\langle \frac{\alpha G^a_{\mu\nu} \widetilde{G}_a^{\mu \nu}}{8 \pi f } \Big\rangle_{\text{non-eq}}^\text{th} = \frac{\Gamma_{\text{sph}}}{2Tf}\frac{\dot{\phi}}{f} \, ,
\end{equation}
where we have used the dissipation-fluctuation theorem to relate the diffusion coefficient of the Chern-Simons number, $\Gamma_{\text{sph}}$, to the linear response induced by $\langle \dot{\phi} \rangle/f$.
Matching with Eq.~\eqref{eq:linear_response} and Eq.~\eqref{eq:EulerLagrangeWarmInflation}, the thermal friction coefficient due to sphaleron heating is 
\begin{equation}
\label{eq:sphaleronrate}
\Upsilon_{\text{sph}}(T) \approx \frac{\Gamma_{\text{sph}}}{2Tf^2} = \kappa \alpha^5 \frac{T^3}{f^2} \, .
\end{equation}

In summary, minimal warm inflation has a unique particle production mechanism exploiting the topology of non-abelian gauge theories, while being protected from mass corrections that spoil the inflaton potential.

\subsection{\edit{Sphaleron heating} in the presence of light quarks}
\label{subsec:MWIwQCD}
Given the advantages of minimal warm inflation as a microphysical theory, the prospects of QCD functioning as the non-abelian gauge group are tantalizing. Sphaleron heating in the presence of fermions has been discussed in \cite{Berghaus:2020ekh} for fermions in a general representation. Here we recap the argument in \cite{Berghaus:2020ekh} using the example of fermions in the fundamental representation to illuminate how the dynamics change for QCD. Let us consider a toy model with $N_c = 3$, and one Dirac fermion $\psi$ in the fundamental representation such that 
\begin{align}
\mathcal{L}_{\text{bath}}  =& - \frac{1}{4} G^a_{\mu \nu }G_a^{\mu \nu} + \bar{\psi}(i \slashed{D} - m) \psi \, ,\nonumber \\
=& - \frac{1}{4} G^a_{\mu \nu }G_a^{\mu \nu} + i \bar{\psi}_L \slashed{D} \psi_L + i\bar{\psi}_R \slashed{D} \psi_R - m \bar{\psi}_L \psi_R - m \bar{\psi}_R \psi_L   \, ,
\end{align}
where we have expanded the fermion terms in chiral right and left-handed fields, $\psi_{R/L} = 1/2 (1 \pm \gamma_5) \psi$, and $ D_u =(\partial _\mu + i g A^a_\mu T^a)$, where $A^a_\mu$ are the gauge bosons of the SU(3), and $T^a$ are the generators of the SU(3) in the fundamental representation. $g$ denotes the gauge coupling $\alpha = g^2/4\pi$. In the massless limit $m \to 0$, the lagrangian exhibits chiral symmetry, e.g. invariance under a chiral phase rotation
\begin{equation}
\psi_L \to e^{i \alpha} \psi_L\,, \, \,\,\,\,\,\, \psi_R \to e^{-i \alpha} \psi_R\,.
\end{equation}
The associated Noether current
$j_A^{\mu} = \bar{\psi} \gamma^\mu \gamma^5\psi$, with charge 
\begin{equation}
j^0_A = \left(\psi_R^\dagger \psi_R - \psi^{\dagger}_L \psi_L \right) = (n_R - n_{\bar{R}}) -(n_L -n_{\bar{L}}) = n_r -n_l \, ,
\end{equation}
is classically conserved. Here, $n_R +n_{\bar{L}} =n_r$, is the net right handed particle number, and, $n_L + n_{\bar{R}} = n_l$, is the net left handed particle number.  
 However, quantum correction lead to an anomalous non-conservation such that 
\begin{equation}
\partial_\mu j_A^\mu = \frac{\alpha G^a_{\mu \nu} \widetilde{G}_a^{\mu \nu}}{4 \pi} \, ,
\end{equation}
which in the presence of sphaleron transitions,
leads to an evolution of $n_r-n_l$,
\begin{equation}
\partial_t (n_r -  n_l) = \left \langle \frac{\alpha}{4\pi}  G^a_{\mu\nu} \widetilde{G}_a^{\mu \nu} \right \rangle \, .
\end{equation}
This non-conservation of chirality leads to a build up of axial charge $j_A^0$ density with an associated chemical potential $\mu_A$, where the Fermi Dirac distribution of each chirality obeys 
\begin{equation}
f_{L,R}(p) = \frac{1}{e^{(p \pm \mu_A)/T}+1} = f_{\bar{R},\bar{L}}(p) \, .
\end{equation}
The axial charge density then relates to the chemical potential as 
\begin{align}
j_A^0 = n_r - n_l &= 2 N_c \int \frac{d^3p}{(2\pi)^3} \left(\frac{1}{e^{(p-\mu_A)/T}+1} -\frac{1}{e^{(p+\mu_A)/T}+1} \right) \, , \nonumber \\ 
 &= \frac{N_c  \mu_A T^2}{3} \, .
\end{align}
For $N_c = 3$, this simplifies to $\mu_A = j_A^0 /T^2$.
The axial chemical potential back-reacts on the biased sphaleron transitions, counteracting the chemical potential of the Chern-Simons number provided by the inflaton, 
\begin{equation}
\left \langle \frac{\alpha G^a_{\mu\nu} \widetilde{G}_a^{\mu \nu}}{8 \pi} \right \rangle_{\text{non-eq}}^{\text{th}} \, = \frac{\Gamma_{\text{sph}}}{2T} \left(\frac{\dot{\phi}}{f} -2 \mu_A \right) \,.
\end{equation}
Thus, the sphaleron dynamics and consequently the particle production via sphalerons from the rolling axion field is drastically altered in the presence of a massless fermion \cite{Berghaus:2020ekh}. 
The coupled evolution of the axial charge and the inflaton in an expanding universe is governed by the coupled set of differential equations, 
\begin{align}
\label{eq:coupled}
 \ddot{\phi} +3 H \dot{\phi} + V' & = - \Upsilon_{\text{sph}} \left(\dot{\phi}  - 2\frac{f j^0_A}{T^2}   \right)\, , \\
 \label{eq:coupled2}
 \dot{j}_A^0 +3H j^{0}_A & =2 \Upsilon_{\text{sph}} f \left(  \dot{\phi} - 2 \frac{f j^0_A}{T^2}\right) \, .
\end{align}
Considering the regime in which Hubble expansion is small compared to $\Upsilon_{\text{sph}}$, and starts with $j_A^0 \neq 0$, we can see that $\dot{\phi}$ efficiently builds up an axial charge with a chemial potential that asymptotes towards canceling the chemical potential of the Chern-Simons number. The coupled  evolution of Eq.~\eqref{eq:coupled} and Eq.~\eqref{eq:coupled2} still allows for particle production, albeit very suppressed, and does not have a slow-roll solution dominated by thermal friction \cite{Berghaus:2019whh}.  

However, SM QCD has quarks with finite masses, which can relax $j_A^0$ via chirality violating scattering processes in the plasma. Naively, one may guess that this rate should scale as $\Gamma_{\text{ch}} \propto m^2 \alpha^2/T $, however it has been shown that infrared divergences enhance the rate by a factor of $\alpha$ \cite{Boyarsky:2020cyk}. Defining the chirality violating rate as 
\begin{equation}
\Gamma_{\text{ch}} = r N_c \alpha \frac{m^2}{T} \, ,
\end{equation}
where $r$ is an $\mathcal{O}(1)$ number, which for an electron photon plasma has been calculated to be $\approx 0.24$ \cite{Boyarsky:2020cyk}, the evolution of the axial charge changes to 
\begin{align}
\label{eq:current}
 \dot{j}_A^0 +3H j^{0}_A & =2 \Upsilon_{\text{sph}} f \left( \dot{\phi} - 2\frac{f j^0_A}{T^2} - \frac{j^0_A \Gamma_{\text{ch}}}{\Upsilon_{\text{sph}}f }\right) \, .
\end{align}
Neglecting the Hubble expansion term and setting $\dot{j}_A^0 = 0$, one can solve Eq.~\eqref{eq:current} for $j_A^0$, and plugging back into Eq.~\eqref{eq:coupled}, one obtains an effective friction coefficient, 
\begin{equation}
\label{eq:eff}
\Upsilon_{\text{eff}} = \Upsilon_{\text{sph}} \left(\frac{\Gamma_\text{ch}}{\Gamma_{\text{ch}} + 2\frac{f^2}{T^2} \Upsilon_{\text{sph}}} \right)\, .
\end{equation}
Eq.~\eqref{eq:eff} has two limiting regimes. When chirality flipping processes are efficient,
\begin{equation}
\label{eq:cond}
 \Gamma_{\text{chir}} \gg 2 \frac{f^2}{T^2} \Upsilon_{\text{sph}} \,, \, \, \,\,\,\,\,\,\,\,\, \text{or} \, \,\,\,\,\,\,\, m \gg \alpha^2 N_c^2 T  \, ,
\end{equation}
sphaleron heating remains the same \cite{Berghaus:2020ekh}. Naively, one would already expect sphaleron heating to remain unchanged for theories with heavy fermions $m \gg f,T$ that can be integrated out and effectively described by a pure Yang-Mills theory. 
Note that the condition in Eq.~\eqref{eq:cond} is much less stringent and allows for fermions to be light enough to be thermalized in the plasma.

In the opposite regime the mass of the lightest fermion suppresses the particle production rate to 
\begin{equation}
 \label{eq:friction_eff}
\Upsilon_{\text{eff}} = \frac{r N_c \alpha }{2} \frac{m^2}{f^2} T \, . 
\end{equation}
Let us consider sample values for QCD. The up quark with $m_u = 3 \, \text{MeV}$ limits the efficiency of particle production. Taking a temperature that can satisfy the weak coupling requirement $\alpha \ll 1$, $T = 100 \, \text{GeV}$, we can get a friction coefficient $\Upsilon_{\text{eff}} \approx 10^{-8} \, \text{GeV}$. To be in a self-consistent warm inflation scenario with QCD one requires $T > H$, and $T \gtrsim 100 \, \text{GeV}$. Additionally, observations place the usual constraints on the spectral tilt $n_s$, number of observed e-folds, $N$, and the amplitude of the power spectrum $A_s $. When attempting to construct a model that can fit observables with the effective QCD friction coefficient in Eq.~\ref{eq:friction_eff}, we find that in the weak regime the former condition is violated, while in the strong regime the latter condition cannot be satisfied. 

Since the light quark masses are what spoil a successful solution for warm inflation with QCD sphalerons, we turn our focus to constructing a model in which quark masses are heavy during inflation, such that sphaleron heating in QCD can proceed efficiently with $\Upsilon_{\text{sph}}$. 
We discuss this idea in detail in Sec.~\ref{sec:MWIQCD}.

\section{Towards warm inflation with QCD sphaleron heating}
\label{sec:MWIQCD}
\subsection{Heavy quarks from additional Higgs vacua}\label{subsec:SMvacua}

Warm inflation with a QCD axion is only possible if the ordinarily light quarks satisfy Eq.~\eqref{eq:cond} during the inflationary phase. 
One can imagine various ways of accomplishing this, such as changing the Yukawa couplings at early times, adding additional mass contributions from BSM physics that become negligible at late times, or by changing the contribution from the SM Higgs.
We will consider the last of these approaches, letting the Higgs sit in a new UV minimum during inflation, making the quarks heavy.
Similar scenarios have been discussed in~\cite{Hook:2019zxa}, although not in the context of warm inflation.
The position of this UV minimum will dictate the required reheating temperature to return to the EW minimum after inflation ends, which in turn will dictate our allowed axion parameter space in Sec.~\ref{sec:warmhybridinflation}.

\begin{figure}
    \centering
    \includegraphics[width=0.8\textwidth]{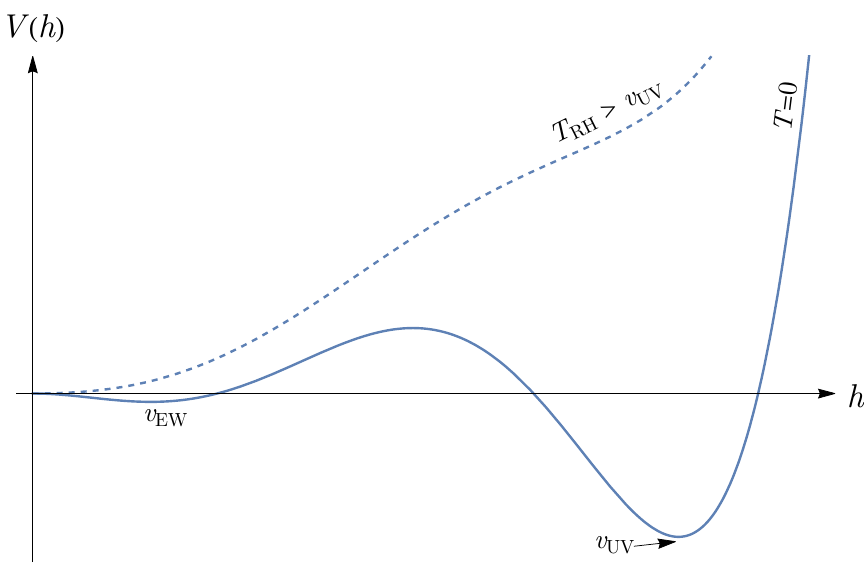}
    \caption{A sketch of the Higgs potential we consider at $T=0$ (solid lines) and after reheating (dashed line). During inflation the Higgs lives in the UV minimum $v_\text{UV}$, making the quarks heavy. After reheating, thermal corrections lift the Higgs out of its UV minimum and allow it to relax to our electroweak minimum $v_\text{EW}$ as the universe cools.}
    \label{fig:sketch}
\end{figure}

The basic ideas of our approach are as follows.
We first assume that the Higgs quartic coupling $\lambda_H$ becomes negative at high scales due to renormalization group running, as it does in the SM.
To turn the resulting unbounded potential into a new UV minimum at $v_\text{UV}$, we will add a higher dimensional operator that becomes important shortly after $\lambda_H$ becomes negative. 
The resulting potential will schematically look like the solid blue curve in Fig.~\ref{fig:sketch}.
\edit{To return to the electroweak minimum after inflation ends, we require} that the reheating temperature after inflation $T_\text{RH}$ is large enough that thermal corrections to the Higgs potential overcome the UV minimum, $T_\text{RH} \gtrsim v_\text{UV}$. 
The resulting potential looks like the dashed blue curve in Fig.~\ref{fig:sketch}. 
The Higgs relaxes to the electroweak minimum $v_\text{EW}$ as the universe cools.\footnote{Due to thermal corrections set by $T_{\text{RH}}$ the Higgs field will start oscillating around $h = 0$. Particle production from the oscillation rapidly damps the amplitude of the Higgs field with a rate \cite{, Hook:2019zxa} $\Gamma_h \sim 10^{-3} T > H$ that exceeds Hubble. Thus, the amplitude of the oscillation decreases much faster than the temperature, ensuring that the Higgs field ends up in the EW-vacuum after reheating \cite{Hook:2019zxa}. }
After the temperature cools below $v_{\text{UV}}$ the UV-vacuum reappears, however, the barrier between the two vacua remains enhanced by thermal contributions. These thermal contributions to the barrier stabilize the EW-vacuum as the universe cools.\footnote{The EW vacuum decay rate at finite temperature in the SM for an unbounded potential has been calculated to be exponentially suppressed below $T \lesssim M_{\text{pl}}$ \cite{DelleRose:2015bpo, Salvio:2016mvj, Markkanen:2018pdo}. In our scenario with a bounded potential the vacuum decay rate is smaller than in the SM, thus vacuum stability is guaranteed.} 
We achieve large reheating temperatures that allow $T_{\text{RH}}$ to exceed the temperature during warm inflation $T$ with a hybrid inflation setup, where a waterfall field provides the energy density that sets $T_{\text{RH}}$, which we discuss further in  Sec.~\ref{sec:warmhybridinflation}. However, in general our idea is compatible with other mechanisms that achieve a large reheating temperature such as for example hilltop inflation.

We use PyR@TE~\cite{Sartore:2020gou} to compute the renormalisation group evolution (RGE) at three-loop order for the gauge couplings and at two-loop order for the Yukawa and Higgs quartic couplings.
Our RGE is initialized at $\mu=m_t$ with relevant inputs given by~\cite{Workman:2022ynf} $m_H = 125.25 \pm 0.17$ GeV, $m_W = 80.377 \pm 0.012$ GeV, $\alpha_S(m_Z) = 0.1180 \pm 0.0009$, and a conservative\footnote{
An additional uncertainty on $m_t$ arises from the interpretation of the top mass parameter $m_t^\text{MC}$ extracted in direct measurements as the pole mass, which differ by non-perturbative effects~\cite{Nason:2017cxd,Hoang:2020iah,Workman:2022ynf}. 
This theoretical uncertainty is estimated to be of order $0.5- 1 \, \text{GeV}$, which could expand our \edit{parameter space to lower reheating temperatures.}
} $m_t = 172.69 \pm 0.3$ GeV from direct measurements, all converted to $\overline{\textrm{MS}}$ inputs using the results of~\cite{Buttazzo:2013uya}. 
For simplicity, we only include the top Yukawa coupling since the others are negligibly small corrections by comparison. 
At high scales, we may neglect the quadratic term in the Higgs potential and write the effective potential as~\cite{Buttazzo:2013uya}
\begin{align}
\begin{split}
    V_\text{eff}^\text{SM} (h)=&\frac{1}{4}\lambda_\text{eff}(h) h^4 \, ,\\
    =&\frac{1}{4}e^{4 \Gamma(h)}\left(\lambda_H(\mu = h) + \lambda^{(1)}_\text{eff}(\mu = h) + \lambda^{(2)}_\text{eff}(\mu = h) + \ldots\right) h^4 \, ,
\end{split}
\end{align}
where $\lambda_H$ is the running Higgs quartic coupling and $\lambda_\text{eff}^{(n)}$ is the $n$-loop correction, which combine into the effective quartic coupling $\lambda_\text{eff}$. The prefactor accounts for the running of the Higgs field itself, with $\Gamma(h) = \int_{m_t}^h \gamma(\mu) d \ln(\mu)$ where $\gamma$ is the anomalous dimension for $h$.
We work up to 2-loop order in the SM effective potential using the results given in Appendix C of~\cite{Buttazzo:2013uya} for $\lambda^{(n)}_\text{eff}$ and $\gamma$.

As written, the effective potential becomes negative shortly after $\lambda_H$ becomes negative. 
To stabilize the potential and yield a new UV minimum\footnote{For some specific fine-tuned values of $\Lambda$, this second minimum may be higher than our electroweak minimum. An investigation of these different regimes and some explicit realisations was recently presented in~\cite{Steingasser:2023ugv}.}, we add the dimension six operator $\mathcal{O}_H$
\begin{equation}
    V_\text{eff} = V_\text{eff}^\text{SM}+\frac{1}{\Lambda^2}(H^\dagger H)^3 \, ,
\end{equation}
where $\Lambda$ is the cutoff scale of the EFT and we have absorbed the associated Wilson coefficient $C_H$ into $\Lambda$ for simplicity.
This operator can easily be UV completed by simple extended scalar sectors, though we will not consider any explicit realizations for simplicity and to remain as general as possible. 
The lowest possible value of $v_\text{UV}$ is determined by choosing the smallest $\Lambda$ that yields a second minimum with a depth $V_\text{eff}(v_\text{UV}) < V_\text{eff}(v_\text{EW})$ such that it is the global minimum.

At finite temperature, the effective potential picks up additional corrections that will lift the Higgs out of the UV minimum.
These are given by\footnote{
Some care must be taken to include the effects of daisy resummation~\cite{Quiros:1999jp,Curtin:2016urg,Senaha:2020mop} for a consistent result. 
We have checked that replacing $m_i^2 \rightarrow m_i^2 + \Pi_i$ with $\Pi_i$ the thermal mass in the full effective potential (Truncated Full Dressing, in the language of~\cite{Curtin:2016urg}) does not meaningfully change our results, and so we neglect it for numerical simplicity.}~\cite{Quiros:1999jp}
\begin{equation}
    \Delta V_\text{th}(h, T) = \frac{T^4}{2\pi^2} \left[\sum_i n_i J_B\left[m^2_i(h)/T^2\right]+\sum_{f} n_f J_F\left[m^2_f(h)/T^2\right]\right] \, ,
\end{equation}
where $i$ sums over bosons, $f$ sums over fermions, and $m_i^2(h)$ is the Higgs-dependent mass.
The functions $J_B$ and $J_F$ are given by
\begin{equation}
    J_{B/F}(y^2) = \int_0^\infty dx \, x^2 \log\left[1 \mp e^{-\sqrt{x^2+y^2}}\right] \, ,
\end{equation}
where the sign is $-(+)$ for $J_B(J_F)$. 
For temperatures $T\gtrsim v_\text{UV}$, the above is well approximated by~\cite{Giudice:2003jh,Espinosa:2015qea,Hook:2019zxa}
\begin{equation}
    \Delta V_\text{th}(h,T) \approx0.06 T^2 h^2 e^{-h/(2\pi T)} \, .
\end{equation}
Numerically, this approximation is quite similar to the full calculation, even for temperatures only slightly larger than $v_\text{UV} / (2\pi)$.
To determine our range of $T_\text{RH}$, we impose the requirement that the only minimum in the full effective potential at $T_\text{RH}$ is at $h=0$.
We find that this gives a result very close to $T_\text{RH} \gtrsim v_\text{UV} / (2\pi)$.

\edit{In the SM,} it has long been appreciated that the scale where $\lambda_H$ becomes negative is highly sensitive to $m_t$~\cite{Degrassi:2012ry}.
For our purposes, this also means that the smallest possible $v_\text{UV}$ (and viable $T_\text{RH}$) is also very dependent on $m_t$.
We consider positive variations\footnote{
Changes in the other input parameters $m_H$, $m_W$, and $\alpha_S$ also impact the position of the minimum, however their effects are subdominant to $m_t$ with current precision and so we do not vary them.
} in $m_t$ in increments of $0.3$ GeV up to 5$\sigma$, conservatively ignoring any additional uncertainty from the $m_t$ interpretation problem~\cite{Nason:2017cxd,Hoang:2020iah,Workman:2022ynf}.
For each value of $m_t$, we compute the effective potential and find the smallest $\Lambda$ that gives a new UV global minimum.
Adding the dimension six operator to the effective potential with this $\Lambda$, we then determine the minimum $T_\text{RH}$ to eliminate the second minimum. 
This procedure gives the $T^\text{min}_\text{RH}$ shown in Tab.~\ref{tab:TR}, which will determine the contours of viable parameter space in Sec.~\ref{sec:warmhybridinflation}.
\begin{table}[h]
    \centering
    \renewcommand{\arraystretch}{1.3}
    \begin{tabular}{|c|c|c|c|c|c|c|}\hline
        $m_t$ & 172.69 & 172.99 $(1\sigma)$ & 173.29 $(2\sigma)$ & 173.59 $(3\sigma)$ & 173.89 $(4\sigma)$& 174.19 $(5\sigma)$\\\hline
        $\Lambda^\text{min}$ & $6.1\times 10^{12}$ & $2.2\times 10^{12}$ & $9.2\times 10^{11}$ & $4.2\times 10^{11}$ & $2.1\times 10^{11}$ & $1.1\times 10^{11}$ \\
        $v_\text{UV}^\text{min}$ & $2.4\times 10^{11}$ & $9.4\times 10^{10} $ & $4.2 \times 10^{10}$ & $2.0\times 10^{10}$ & $1.0\times 10^{10}$& $ 5.6\times 10^9$ \\
        $T_\text{RH}^\text{min}$ & $3.4\times 10^{10}$ & $1.4\times 10^{10}$ & $6.3\times 10^9$ & $3.1\times 10^9$ & $1.7\times 10^9$ & $9.3\times 10^8$ \\ \hline
    \end{tabular}
    \caption{The minimum $\Lambda^\text{min}$ to yield a new UV global minimum, its corresponding position $v^\text{min}_\text{UV}$, and the minimum reheating temperature $T_\text{RH}^\text{min}$ for thermal corrections to bring the Higgs back to our electroweak minimum after inflation ends. We conservatively show positive variations in $m_t$ of $0.3$ GeV~\cite{Workman:2022ynf} up to $5\sigma$, ignoring any additional uncertainty from the $m_t$ interpretation problem~\cite{Nason:2017cxd,Hoang:2020iah,Workman:2022ynf}. All values are in GeV.}
    \label{tab:TR}
\end{table}

\edit{Of course, any deviation from the SM running of $\lambda_H$ could have significant impact on $v^\text{min}_\text{UV}$ and resulting $T^\text{min}_\text{RH}$. 
In Appendix~\ref{subsubsec:lowervacua}, we provide an explicit example showing how the lower $v_\text{UV} \lesssim 10^9$ GeV region may be populated by the presence of additional heavy fermions.
Such a scenario is by no means necessary for our mechanism to work, although the resulting parameter space would be easier to probe with current and future experiments, as demonstrated in Sec.~\ref{sec:constraints}.
}

\subsection{The heavy QCD axion}\label{sec:heavyQCDaxions}

A heavy axion, characterized by UV contributions to its potential alongside the instanton mass resulting from the coupling that triggers QCD sphaleron heating, is an essential condition for achieving minimal warm inflation with the axion as the inflaton \cite{Berghaus:2019whh}. 
Thus, achieving warm inflation via coupling the QCD axion to QCD gluons requires a heavier axion than the standard QCD axion prediction. 

The QCD axion has seen decades of attention as an elegant solution to the strong CP problem~\cite{Weinberg:1977ma,Peccei:1977hh,Wilczek:1977pj,Peccei:1977ur} and potential dark matter candidate~\cite{Preskill:1982cy,Dine:1982ah,Abbott:1982af,Adams:2022pbo}. 
There exist a number of excellent reviews on axions in various contexts, see for example~\cite{Marsh:2015xka,DiLuzio:2020wdo,Hook:2018dlk,Adams:2022pbo}. 
Here we will give a very brief review of the QCD axion and its heavy variations.
The typical QCD axion begins as a massless periodic pseudoscalar $\phi \cong \phi + 2\pi f$ coupled to $SU(3)_C$ with a Lagrangian\footnote{
Note that we have absorbed the quantized coefficient $N$ in the axion coupling to gluons into the definition of $f$. 
We have also omitted any couplings to the $SU(2)_L$ and $U(1)_Y$ gauge fields, which may generically be necessary depending on the global structure of the gauge group and the value of $N$~\cite{Choi:2023pdp,Reece:2023iqn,Agrawal:2023sbp,Cordova:2023her}. 
The presence of these couplings would not change the inflationary behaviour we are interested in, although they are important for some of the projected constraints presented in Sec.~\ref{sec:constraints}. }
\begin{equation}\label{QCDAxionLagr}
    \mathcal{L}\supset \frac{1}{2}\partial_\mu \phi \partial^\mu \phi + \frac{\alpha_S}{16\pi f}\phi  \epsilon^{\mu\nu\alpha\beta} G_{\alpha\beta}^a G_{\mu\nu}^a \, ,
\end{equation}
where $\alpha_S$ is the strong coupling constant and $f$ is the axion decay constant.
After confinement, QCD dynamics give a potential to the axion which is minimized at the CP-preserving value $\phi = 0$ by the Vafa-Witten theorem~\cite{Vafa:1984xg}, 
allowing it to solve the strong CP problem as it relaxes. 
One can find this potential explicitly by matching onto the chiral Lagrangian,
\begin{equation}
    V_{\textrm{QCD}}(\phi) = - m_\pi^2 f_\pi^2 \sqrt{1-\frac{4 m_u m_d}{(m_u+m_d)^2}\sin^2\frac{\phi}{2f}} \, ,
\end{equation}
where $m_\pi$ and $f_\pi$ are the pion mass and decay constant, respectively, and $m_{u,d}$ are the up and down quark masses. 
Alternatively, a semiclassical analysis in the dilute-instanton-gas approximation yields a potential $V(\phi) = -m_\phi^2 f^2 \cos(\phi/f)$. 
The forms of the two expressions correspond in the small $\phi$ limit, but only the former is valid for larger field values at zero temperature. 
At temperatures above the chiral phase transition, the instanton calculation becomes reliable, and the potential as a whole becomes suppressed. 
In either case, the induced potential breaks the shift symmetry of the axion and gives a relation between the mass $m_\phi$ and decay constant $f$, leading to the usual QCD line on axion constraint plots. 

A common UV completion for the Lagrangian in Eq.~\eqref{QCDAxionLagr} is a theory with a $U(1)_{PQ}$ Peccei-Quinn symmetry under which chiral fermions and at least one complex scalar are charged. 
The axion is the pseudo-Nambu-Goldstone boson of this broken symmetry and inherits the coupling to the gauge fields from the Adler-Bell-Jackiw (ABJ) anomaly between $U(1)_{PQ}$ and $SU(3)_C$.
If the axion comes from such a symmetry, there is a potential problem known as the axion quality problem~\cite{Georgi:1981pu,Dine:1986bg,Ghigna:1992iv,Barr:1992qq,Kamionkowski:1992mf,Holman:1992us,Randall:1992ut}. 
Since $U(1)_{PQ}$ has an ABJ anomaly, it is not a good symmetry of the quantum theory in the first place, and so there is no reason why it should not be broken at higher scales.
In fact, even in the absence of this anomaly, quantum gravity is generally thought to gauge or break all global symmetries (see for example~\cite{Reece:2023iqn} and references therein), so one generically expects to see terms such as 
\begin{equation}
    \mathcal{L} \supset \frac{\Phi^n}{M_\text{Pl}^{n-4}} \supset \frac{f^n}{M_\text{Pl}^{n-4}} \cos\left(\frac{\phi}{f}+\theta_n \right) \, ,
\end{equation}
where $\Phi\sim f e^{i \phi/f}$ is the scalar charged under $U(1)_{PQ}$ that gives rise to the axion after obtaining a nonzero vacuum expectation value, and $\theta_n$ is a nonzero phase which generically causes the minimum of the axion potential to deviate from $\theta_{\textrm{QCD}}=0$. 
In order to solve the strong CP problem within experimental precision, $|\theta_{\textrm{QCD}}| \lesssim 10^{-10}$~\cite{Abel:2020pzs}, the potential from QCD must dominate, $V_\textrm{QCD} \gtrsim f^n/M^{n-4}_{Pl}$. 
For typically considered values of $f$, this translates to a stringent constraint on $n$.
For example, if we consider $f\sim 10^{12}$ GeV with $\theta_n \sim \mathcal{O}(1)$, this bound becomes $n \geq 14$.
Forbidding operators of such high mass dimension is generally quite challenging.
Resolving this has been one of the primary focuses of axion modelbuilding in recent years. 

For our purposes, we will need a significantly larger potential than the simplest QCD axion can achieve, since the QCD induced potential is suppressed at nonzero temperatures.
We therefore must consider mechanisms that add to the axion mass while preserving the solution to the strong CP problem.
However, note that this modelbuilding direction is motivated by solutions to the quality problem as well. 
To see this, consider adding another contribution to the potential of the generic form
\begin{equation}
    V_\text{BSM}(\phi) = -\Lambda^4 \cos{\left(\frac{\phi}{f}\right)} \, ,
\end{equation}
where $V_\text{BSM}$ is also minimized when $\phi=0$, and $\Lambda \gg \Lambda_\text{QCD}$. 
This potential then generates a larger axion mass and reduces the quality problem simultaneously, since the condition $V_\text{BSM} > f^n/M^{n-4}_{Pl}$ is easier to satisfy. 
The modelbuilding challenge is to generate such a potential that is also minimized at $\theta = 0$, i.e. it also solves the strong CP problem.

Additional contributions to the axion potential of this form can be generated in a number of ways. 
One method involves introducing enough new heavy fields charged under $SU(3)_C$ that $\alpha_S$ runs to strong coupling again at high energies~\cite{HOLDOM1982397, HOLDOM1985316,FLYNN1987731,Kitano:2021fdl}. 
This makes QCD instanton contributions greatly enhanced over the SM expectation.
This may also be done by introducing an extra dimension at high energies~\cite{Gherghetta:2020keg} which leads to enhanced instanton contributions. 
Similarly, embedding $SU(3)_C$ into a UV theory can lead to a larger axion mass through UV dynamics~\cite{DIMOPOULOS1979435,Gaillard:2018xgk,Csaki:2019vte,Kivel:2022emq,Gherghetta:2016fhp}.
Alternatively, one can couple the axion to extra dark gauge groups~\cite{PhysRevLett.47.1035,Rubakov:1997vp,Berezhiani:2000gh,Hook:2014cda,Fukuda:2015ana,Dimopoulos:2016lvn,Hook:2019qoh,Dunsky:2023ucb,Co:2022bqq,Valenti:2022tsc}, with a symmetry that imposes $\theta_\textrm{QCD}=0$ to minimize the total potential.
One way of realising this is to additionally couple the axion to a heavy mirror sector obeying a $\mathbb{Z}_2$ symmetry such that the minimum of the potential induced by the mirror sector is identical to our own, $\theta_\textrm{QCD}=\theta_\textrm{Mirror}=0$.
Another solution~\cite{Agrawal:2017ksf} involves introducing multiple axions, each coupling to a new strongly coupled $SU(3)$ gauge group from which our $SU(3)_C$ is the diagonal subgroup.
This can lead to a wide variety of axion mass spectra depending on the number of $SU(3)$ factors and size of their respective gauge couplings, including regimes where all axion masses are much heavier than the generic prediction. 

As an explicit example of a model that gives some of the mass range we explore, consider the $SU(3)^3$ model of~\cite{Agrawal:2017ksf}. 
In their benchmark 2, they find two axions with masses of $\mathcal{O(\textrm{GeV})}$ with values of $f\sim 10^{1}-10^9$ GeV, overlapping with our viable parameter space at the price of some mild tension with experimental constraints on the lighter axion.
Variants with more $SU(3)$ factors would easily enable heavier axions in our range of parameters while avoiding these constraints altogether.
For our purposes, we remain agnostic to the details of the UV dynamics making our axion heavy and consider $f$ and $m$ to be independent parameters.
For the interested reader, the review~\cite{DiLuzio:2020wdo} includes a more comprehensive summary of various modelbuilding efforts to break the usual $m_\phi-f$ relation induced by QCD. In summary, the heavy QCD axion is well motivated by the quality problem beyond its role as the warm inflaton in this work. 
 
\section{\edit{An example: hybrid warm inflation with QCD sphaleron heating}}\label{sec:warmhybridinflation}
Having set the stage for how heavy quarks during warm inflation restore efficient sphaleron heating $\Upsilon_{\text{sph}}$, we now construct an explicit warm inflation model that matches cosmological observables, and enables the Higgs field to relax to its EW vacuum after inflation. The key ingredient that allows thermal corrections to the Higgs potential to lift the Higgs field out of its UV-vacuum is a hierachy in temperatures. In particular, the temperature during warm inflation $T$ has to be smaller than $v_{\text{UV}}$, while the reheating temperature after inflation $T_\text{RH}$ has to be larger. This hierachy, $T \ll v_{\text{UV}} \lesssim T_\text{RH}$, cannot be achieved in single field warm inflation with a monomial potential, as the temperatures evolves by at most an $\mathcal{O}(1)$ factor, leading to $T_\text{RH} \approx T$. \edit{However, as briefly mentioned in Section~\ref{sec:overview} due
to the predicted spectral tilt in the strong regime of warm inflation with sphaleron heating, 
renormalizable monomial potentials are already ruled out by observations. Fine-tuned potentials such as $V \propto \phi^5$ are able to match the observed spectral tilt, but are not compelling. A hybrid inflation setup is a natural candidate to achieve the required hierarchy in slow-roll parameters $\eta_V > 2\epsilon_V$, that can accommodate the observed spectral tilt without tuning \cite{Berghaus:2019whh}. Moreover, }hybrid inflation naturally has a separation of scales, since the reheating temperature is determined by the energy density of the waterfall field $\sigma$, which can exceed the temperature during warm inflation, $T$, by many orders of magnitude. \edit{Thus, hybrid warm inflation in our model achieves two things at the same time. It allows for a viable warm inflationary scenario without a fine-tuned potential, and automatically provides the large reheating temperature necessary in order for the Higgs to relax to its EW-vacuum.}

\subsection{The scalar power spectrum and spectral tilt}
We begin with briefly reviewing the predictions for the amplitude of the scalar power spectrum and the spectral tilt in warm inflation with $\Upsilon_{\text{sph}}$, and then proceed to outline the viable parameter space for hybrid warm inflation with the heavy QCD axion. Several analysis calculating the scalar power spectrum in warm inflation have been performed. The first one including temperature dependent friction coefficients \cite{2009JCAP...07..013G} performed by Graham and Moss from 2009 applicable for sphaleron heating with $\Upsilon_{\text{sph}} \propto T^3$, identified a growing mode in the inflaton perturbations that enhances the power spectrum by $\sim Q^9$. Bastero-Gil, Berera, and Ramos \cite{Bastero-Gil:2011rva} in 2011 also included temperature dependence and found a growing mode, however motivated by their choice of model they considered a thermal friction scaling of $\Upsilon \propto T^3/\phi^2$, thus their results are not directly comparable. Das and Ramos \cite{Das_2020} in 2020 compute a prediction applicable for sphaleron heating, and in independent work in 2022 Mirbabayi and Gruzinov \cite{Mirbabayi:2022cbt} specifically perform an analysis for sphaleron heating.
We find that in our regime of interest, $Q \gtrsim 10$, the differences  have negligible impact on the viable parameter space of hybrid warm inflation with the QCD axion. 
We center our discussion and results on the most recent analysis \cite{Mirbabayi:2022cbt}, but contrast it with the previous results from \cite{2009JCAP...07..013G, Das_2020} when relevant.

Writing the scalar power spectrum in terms of its amplitude $A_s$ and spectral tilt $n_s$, we have
\begin{equation}
\Delta_s^2 (k) \equiv \frac{k^3}{2 \pi^2} P_{\mathcal{R}}(k) =  A_s(k_*) \left( \frac{k}{k_*} \right)^{n_s(k_*)-1} \, ,
\end{equation}
where the CMB measurement for the amplitude $A_s$ in the base $\Lambda$CDM model combining Planck temperature, polarization, cross-correlations, and lensing is  $\ln \left(10^{10} A_s \right) = 3.044 \pm 0.014$ at the pivot scale $k_* = 0.05 \, \text{Mpc}^{-1}$, 
while the spectral tilt is determined to be $n_s = 0.9626 \pm 0.0057$ \cite{Planck:2018jri}.

\begin{figure}
    \centering
    \includegraphics[width=0.49\textwidth]{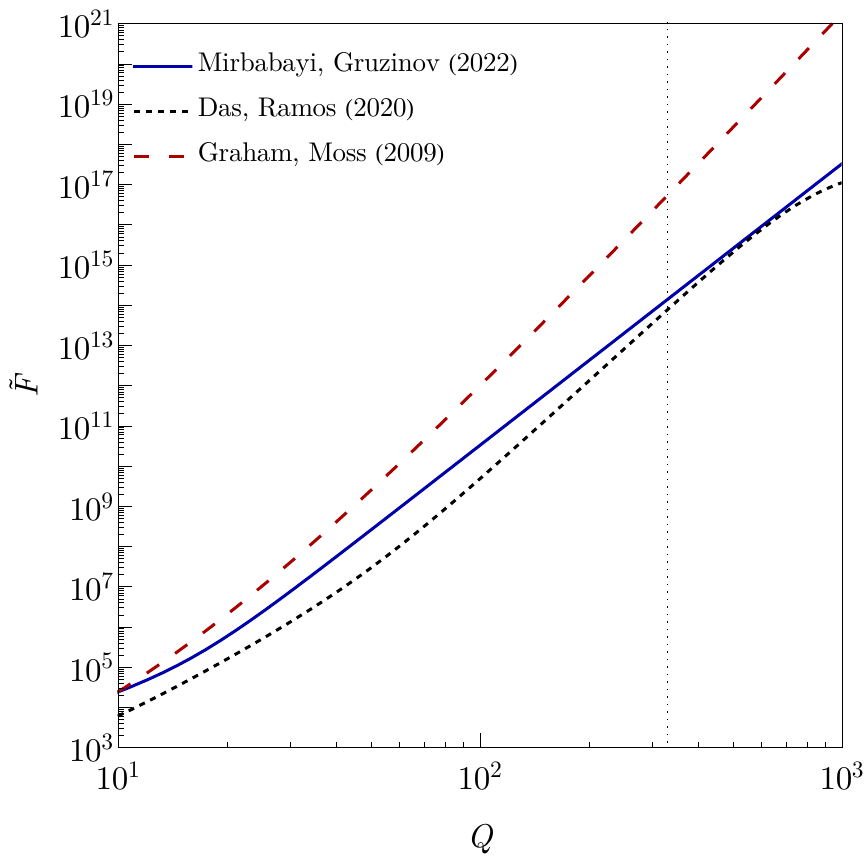}
    \includegraphics[width=0.48\textwidth]{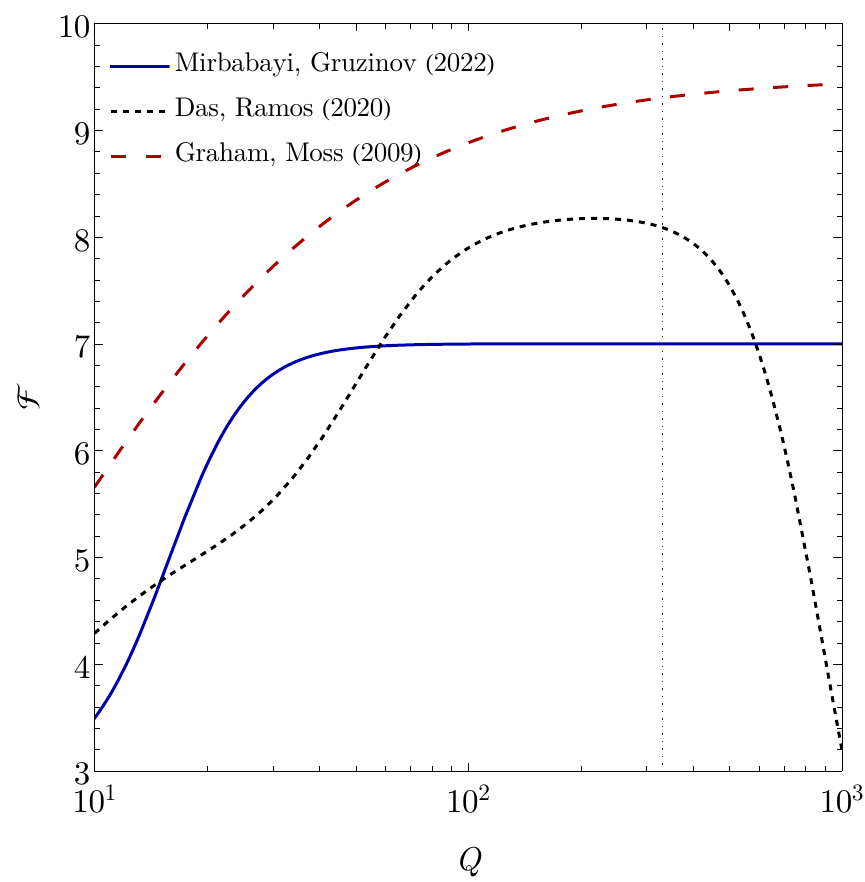}
    \caption{ The differences in the prediction of the scalar power spectrum (\textit{left}) and spectral tilt (\textit{right}) for warm inflation with sphaleron heating from the three applicable results in the literature \cite{Mirbabayi:2022cbt, Das_2020, 2009JCAP...07..013G}. $\tilde{F}(Q)$ is defined by Eq.~\eqref{PowerSpectrumWarmInflation}, and $\mathcal{F}(Q) \equiv  {\text{d}\ln(\tilde{F}(Q))}/\text{d}\ln(Q)$.
    We plot up to $Q = 1000$, but note that the power spectrum in Mirbabayi and Gruzinov \cite{Mirbabayi:2022cbt} is numerically valid up to $Q \approx 330$, which we denote by a vertical dashed line.}
    \label{fig:fderfunc}
\end{figure}

The scalar power spectrum prediction of warm inflation with sphaleron heating $\Upsilon_{\text{sph}} \propto T^3$ from Mirbabayi and Gruzinov \cite{Mirbabayi:2022cbt} is 
\begin{align}
     \Delta_s^2 &\approx \frac{1}{4 \pi^2} \frac{H^3 T}{\dot{\phi}^2} \tilde{F}_M(Q) + \frac{1}{4 \pi^2} \frac{H^4}{\dot{\phi}^2} \, , 
\label{PowerSpectrumWarmInflation}
\end{align}
where\footnote{Note that our function $\tilde{F}_M(Q)$ differs from $F_2(x)$ in \cite{Mirbabayi:2022cbt} since we absorb the extra factor of $3Q$ into the enhancement function, which in \cite{Mirbabayi:2022cbt} uses $\Upsilon/H = 3Q$ as an input.}  
\begin{align}
    \tilde{F}_M(Q) \equiv 0.00032 Q^7 + 168Q \left( \frac{1}{3} \left( 1 + \frac{9Q^2}{25} \right) + \frac{2}{3} \tanh \left( \frac{1}{30Q} \right) \right) \, ,  \label{Ftilde-functionform-Mirbabayi}
\end{align}
which is numerically valid up 
to $Q \approx 330$, with $Q \equiv \Upsilon/3H$ \cite{Mirbabayi:2022cbt}. Focusing on the strong regime $Q \gtrsim 10$, where the second term in Eq.~\eqref{PowerSpectrumWarmInflation} is negligible,
the comparable function quantifying the power spectrum from Das and Ramos \cite{Das_2020} 
is
given by 
\begin{align}
    \tilde{F}_D(Q) = \frac{2 \pi Q \sqrt{3}}{\sqrt{3 + 4 \pi Q}} \left( \frac{1 + 6.12 Q^{2.73}}{(1 + 6.96Q^{0.78})^{0.72}} + \frac{0.01Q^{4.61}(1+4.82 \times 10^{-6} Q^{3.12})}{(1 + 6.83 \times 10^{-13} Q^{4.12})^2} \right) \, . \label{Ftilde-functionform-Ramos}
\end{align}
For completeness also comparing to the Graham and Moss result from 2009, we find \cite{2009JCAP...07..013G}

\begin{align}
    \tilde{F}_{G}(Q) = \sqrt{3 \pi (1 + Q)} \left(1 + \frac{Q}{Q_c} \right)^9 \, , \label{Ftilde-functionform-Graham}
\end{align}

\noindent with $Q_c = 7.3$ \cite{2009JCAP...07..013G}.
We contrast the expressions in Eq.~\eqref{Ftilde-functionform-Mirbabayi}, Eq.~\eqref{Ftilde-functionform-Ramos}, and Eq,~\eqref{Ftilde-functionform-Graham} in Fig.~\ref{fig:fderfunc}. Both recent results predict a more shallow growth of the scalar power spectrum with higher $Q$ compared to Graham and Moss, and agree within an order of magnitude. 
The impact of the differences on the spectral tilt captured to the left of Fig.~\ref{fig:fderfunc} in $\mathcal{F}$ is $\mathcal{O}(1)$. For a comparison of results in the literature also including other models of warm inflation, see \cite{Montefalcone:2023pvh}.  

The warm inflation temperatures are well below Planck scales, and thus the tensor perturbations are the same as in cold inflation \cite{2000cils.book.....L}
\begin{align}
    \Delta_h^2 = \frac{2}{\pi^2} \frac{H^2}{M_{\text{Pl}}^2} \, ,
\end{align}

\noindent resulting in a strongly suppressed tensor-to-scalar ratio $r \equiv \frac{\Delta_h^2}{\Delta_s^2} \approx 0$ due to the relative enhancement by factors of $T/H$, and $Q$. This suppression could be an explanation for the non-observation of tensor modes thus far. 
Calculating the spectral tilt from Eq.~\ref{PowerSpectrumWarmInflation}, we find 
\begin{align}
    n_s - 1 &= \frac{d \ln(\Delta_R^2)}{dN} = 3 \dv{\ln(H)}{N} + \dv{\ln(T)}{N} - 2 \dv{\ln(\vert \dot{\phi} \vert)}{N} + Q \dv{\ln(\tilde{F}(Q))}{Q} \dv{\ln(Q)}{N} \, , \nonumber \\
    &= \left( \frac{10}{7} \frac{Q \tilde{F}'(Q)}{\tilde{F}(Q)} - 2 \right) \varepsilon_V - \frac{6 Q \tilde{F}'(Q)}{7 \tilde{F}(Q)} \eta_V \, , \nonumber \\
    &\equiv \left( \frac{10}{7} \mathcal{F}(Q) - 2 \right) \varepsilon_V - \frac{6}{7}\mathcal{F}(Q) \eta_V \, , \label{ScalarTiltWarmInflation}
\end{align}
where we have introduced the function $\mathcal{F}(Q) \equiv \dv{\ln(\tilde{F}(Q))}{\ln(Q)}$, plotted in Fig.~\ref{fig:fderfunc} alongside various enhancement functions $\tilde{F}(Q)$. For  $Q > 50$, $\mathcal{F}_M(Q) \equiv \dv{\ln(\tilde{F}_M(Q))}{\ln(Q)}$ asymptotes to $7$, such that the spectral tilt simplifies to 
\begin{equation}
n_s - 1 \underrel{Q > 50}{=}  8 \epsilon_V - 6 \eta_V \, .
\end{equation}

As briefly mentioned in the beginning of this section, one can see explicitly here that since we observe a red-tilted spectrum with $n_s \approx 0.963$, we must have that $\eta_V > \frac{4}{3}\varepsilon_V$, a hierachy achievable in hybrid warm inflation without tuning \cite{Berghaus:2019whh}. 

\subsection{Hybrid warm inflation}
Our mechanism of hybrid inflation \cite{Linde:1993cn} involves a coupling between the axion inflaton field, $\phi$, and a complex scalar waterfall field, $\sigma$, with the potential
\begin{align}
    V(\sigma,\phi) = \frac{1}{4\lambda}(M^2 - \lambda |\sigma|^2)^2 + \frac{1}{2}g(\sigma)^2 |\sigma|^2 \phi^2 + \frac{1}{2}m_{\phi}^2 \phi^2 \, . \label{HybridInflationPotential}
\end{align}
Note that while we have written $V(\sigma,\phi)$ in terms of $\phi^2$, since $\phi$ is a compact scalar, any $\phi^2$ terms are implicitly periodic functions such as $\cos (\phi / f_D)$ that have been expanded out for ease of discussion.\footnote{
Here we also assume that $f_D$ is a different, larger scale than $f$ that enables us to inflate on a potential that looks approximately like $\frac{1}{2}m_\phi^2 \phi^2$ for field ranges larger than $f$. Such a hierarchy of scales is possible in, for example, clockwork models~\cite{Kaplan:2015fuy, Choi:2015fiu}.}~The waterfall field, once $\phi$ crosses a critical value $\phi_c = M/g(\sigma)$, undergoes a phase transition and rolls down to its true minimum at $\langle \sigma\rangle = M/\sqrt{\lambda}$, which terminates inflation.
We take the coupling between the axion inflaton field and the waterfall field to depend on $\sigma$ such that $g(0) = g$ during inflation, and $g(\sim M/ \sqrt{\lambda}) = 0$. This dependence is not necessary to achieve minimal warm inflation via QCD sphaleron heating. However, in its absence the QCD axion receives a heavy mass contribution after inflation, proportional to the
vev of the waterfall field $\langle \sigma \rangle = M/\sqrt{\lambda}$, which generically spoils the axion as a solution to the strong CP problem,
and makes it too heavy to be collider accessible.
We therefore propose a setup in which this contribution vanishes in the late universe, an arguably more exciting scenario. We base the motivation for the field dependence $g(\sigma)$ in the coupling
between the inflaton and the waterfall field on a simple UV completion, which features a hidden SU(N) gauge group with hidden quarks charged under a global $U(1)_{\sigma}$~\cite{Gong:2021zem}. We provide a discussion of the UV completion in Appendix~\ref{sec:UV}.

The effective potential prior to the first-order phase transition is
\begin{align}
\label{eq:vacuumdomination}
    V_{\text{eff}}(\phi) = \frac{1}{4\lambda} M^4 + \frac{1}{2} m_{\phi}^2 \phi^2 \gg \frac{1}{2}m_{\phi}^2 \phi^2\, , \,\,\,\,\,\,\,\,\,\,\, \,\,\,\,\,\,\,(\epsilon_V \ll \eta_V)
\end{align}
 ensuring slow rolling for large values of $\phi$ (by our slow-roll parameters $\varepsilon_V, \eta_V \ll 1$), guaranteeing a period of accelerated expansion when $\phi > \phi_c$. 
 We enter the phase transition when $\phi < M/g$, as in this case the field $\sigma$ gains a tachyonic effective mass term, $M_\sigma(\phi)^2 = -M^2 + g^2 \phi^2 < 0$ for $\phi < M/g$.

To ensure instant reheating, 
we demand that after the axion field passes the critical value, it rolls to its minimum ($\phi = 0$) much faster than a Hubble time, $\Delta t = \frac{1}{H}$, which is satisfied if $\vert M_\sigma(\phi) \vert ^2 \gg H(t)^2$. At this minimum it undergoes rapid oscillations, losing energy to the expanding universe \cite{Linde:1993cn}. The oscillations around the true minimum convert the vacuum energy stored in $\sigma$, $\frac{M^4}{4\lambda}$, to radiation with temperature $T_\text{RH}$,
\begin{equation}
T_\text{RH} = \left(\frac{15}{2 \pi^2 g_* \lambda} \right)^{\frac{1}{4}} M \, ,
\end{equation}
 where we use the conservative estimate of $g_* = 106.75$, corresponding to all SM degrees of freedom\footnote{The initial reheating temperature is determined by fewer degrees of freedom since the Higgs field resides in the UV-vacuum, and quarks and gauge bosons do not yet contribute to the relativistic degrees of freedom. Accounting for this transition increases $T_{\text{RH}}$ at most by a factor of two. }.
The axion on the other hand will roll to its minimum at $\phi = 0$, and become irrelevant much faster than a Hubble time as long as \cite{Berghaus:2019whh}

\begin{equation}
    M^3 \ll \frac{\sqrt{\lambda} g m_{\phi} M_{\text{Pl}}^2}{1 + Q_{\text{eff}}} \, , \label{MgConstraintEq1} \,\,\,\,\,\,\,\, \,\,\,\,\,\,\,\,\,\text{(instant reheating)}
\end{equation}

\noindent where $Q_{\text{eff}} \ll 1$ is the dissipative factor during reheating, after the initial phase transition. At that point $T_\text{RH} > v_{\text{UV}}$, thus the quark masses are light again and sphaleron heating becomes suppressed due to a build up of chiral charge leading to $\Upsilon_{\text{sph}} \to \Upsilon_{\text{eff}} \ll \Upsilon_{\text{sph}}$.
Thus, inflation ends within a Hubble time after the phase transition induced from the axion field passing the critical value $\phi_c \equiv M/g$.\footnote{
An additional requirement comes from requiring that the sum of the Higgs potential at the UV minimum and the inflaton potential is positive, $V_\text{eff}(\phi) + V_\text{Higgs}(h) > 0$, to ensure that the Higgs does not alter inflationary dynamics~\cite{Hook:2019zxa}. This condition turns out to be strictly weaker than our other constraints, so we will not discuss it further.}

Using Eq.~\eqref{RadiationEnergyDensityWI}, 
\eqref{eq:SlowRollApprox1} - \eqref{eq:SlowRollApprox3},
we may express the functions $T$ and $Q$
in terms of the hybrid inflation theory parameters. We find

\begin{align}
    T(\phi) &= 
     \left( \left( \frac{f^2}{\kappa \alpha_S^5} \right) \frac{15 M_{\text{Pl}} \sqrt{3}}{\pi^2 g_*} \frac{m_{\phi}^4 \sqrt{\lambda} \phi^2}{M^2} \right)^{1/7} \, , 
    \label{CMBTemperatureFunction} \\
    Q(\phi) &\equiv \frac{\Upsilon_{\text{sph}}}{3H} = 
    \left( \left( \frac{\kappa \alpha_S^5}{f^2} \right)^4 \frac{4.8 \times 10^{4} M_{\text{Pl}}^{10}}{\pi^6 g_*^3} \frac{\lambda^5 m_{\phi}^{12} \phi^6}{M^{20}} \right)^{1/7} \, .
    \label{CMBHeatFunction}
\end{align}

\begin{figure}
    \centering 
    \includegraphics[width=\textwidth]{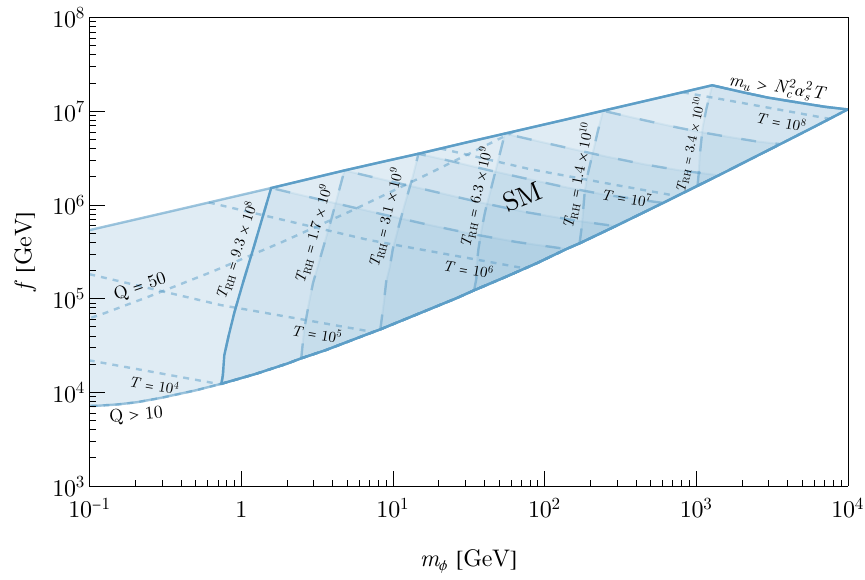}
    \caption{The viable region in the $f-m_\phi$ plane for a heavy QCD axion compatible with being the warm inflaton in hybrid minimal warm inflation given the power spectrum in \cite{Mirbabayi:2022cbt}.
    We display contours of constant $Q$ values, constant $T$ values for temperature during inflation, and reheating temperatures $T_{\text{RH}}$, where the temperatures are in units of GeV.
    Triangular regions indicated by thick dashed lines defined for specific $T_{\text{RH}}$ values (and corresponding UV vacua for the Higgs) arise due to the constraint of Eq.~\ref{eq:cond}, $m_u (T_\text{RH}) > N^2_c \alpha_s^2 T$. 
    The chosen benchmark $T_\text{RH}$ values correspond to conservative $1\sigma$ positive variations in the measurement of the top-quark mass $m_t$ (c.f. Table~\ref{tab:TR}), ignoring any additional uncertainty from the $m_t$ interpretation problem~\cite{Nason:2017cxd,Hoang:2020iah,Workman:2022ynf}. 
    \edit{Letting additional BSM degrees of freedom modify the running of the Higgs quartic self coupling extends the viable region down to $m_\phi\gtrsim 0.1$ GeV, based on the simple vector-like fermion example discussed in Appendix~\ref{subsubsec:lowervacua}.}
    }
\label{fig:parameterspace}
\end{figure}

Now let us consider the inflationary observables discussed at the beginning of this section and see how they map onto the theoretical parameter space in hybrid minimal warm inflation with the heavy QCD axion. 
The first observable to consider is the red tilted scalar tilt ($n_s < 1$). We note that
\begin{align}
    \frac{\varepsilon_V(\phi)}{\eta_V(\phi)} = \frac{1}{2} \left( \frac{4 \lambda m_{\phi}^2 \phi}{M^4} \right)^2 \frac{M^4}{4 \lambda m_{\phi}^2} = \frac{\frac{1}{2}m_{\phi}^2 \phi^2}{\frac{1}{4\lambda}M^4} \ll 1 \, ,
\end{align}
which allows us to 
neglect $\varepsilon_V(\phi)$, such that Eq.~\eqref{ScalarTiltWarmInflation} simplifies to
\begin{align}
    \eta_V(\phi_*) &= \frac{7 (1 - n_s)}{6 \mathcal{F}(Q_*)} \, ,
    \label{eq:eta_ns}
\end{align}
where we have defined $Q_{*} \equiv Q(\phi_*)$.
Using $Q(\phi) \propto \phi^{6/7}$ and defining $\phi_* \equiv \phi_c (1 + \Delta)$,
we relate the scalar tilt $n_s$ to the amount of observed e-folds $N_* \approx 60$ with  Eq.~\eqref{eq:NFoldsWarmInflation}, 

\begin{align}
    N_* &= \frac{1}{M_{\text{Pl}}^2} \int_{\phi_c}^{\phi_*} \frac{M^4}{4 \lambda m_{\phi}^2 \phi} Q(\phi) d\phi \, , \nonumber \\
    &= \frac{7}{6 \eta_V(\phi_*)} (1 - (1 + \Delta)^{-6/7}) \, , 
\end{align}
where expanding to first order in $\Delta$ and plugging in Eq.~\eqref{eq:eta_ns} we find
\begin{equation}
\Delta \approx \frac{7N_*(1 - n_s)}{6\mathcal{F}_M(Q_*)}\underrel{Q > 50}{\approx} \frac{1}{6}N_{*}(1 - n_s)\, \approx 0.37 \, .
\end{equation}
We have six theory parameters in total that characterize the parameter space: $\Delta$, $M$, $g$, $f$, $m_{\phi}$, and $\lambda$, which are constrained by the three observational constraints $N_*$, $n_s$, and $A_s$, self-consistency relations regarding the assumptions we have made in the description of hybrid inflation, and validity of the EFT. Choosing to write the remaining theory parameters in terms of linear combinations of $M/g$, $g$, and $Q$, systematically solving and substituting yields

\begin{align}
    \begin{split}
    f \approx 8.3 &\times 10^{-7} \left( \frac{\kappa \alpha_S^5}{10^{-4}} \right)^{1/2} \left(\frac{Q}{10} \right)^{-\frac{1}{6}}\left( \frac{\tilde{F}(Q)}{2.4 \times 10^4} \right)^{-\frac{1}{6}} \left( \frac{\mathcal{F}(Q)}{3.5} \right) \frac{M}{g} \, , \label{ACDMWIHybridInflation} \end{split} \\
    m_{\phi} \approx 4.2 &\times 10^{-10} \left(\frac{Q}{10} \right)^{\frac{1}{3}}\left( \frac{\tilde{F}(Q)}{2.4 \times 10^4} \right)^{-\frac{2}{3}} \left( \frac{\mathcal{F}(Q)}{3.5} \right)^{-\frac{3}{2}} \frac{M}{g} \, ,
    \label{AxionMassMWIHybridInflation} 
\end{align}
where we use canonical values of $N_* = 55$, $A_s = 2.1 \times 10^{-9}$, and $n_s = 0.963$.
Here $M/g$ is bounded from above by the conditions in Eq.~\eqref{eq:vacuumdomination} and Eq.~\eqref{MgConstraintEq1} which roughly constrain $M/g \ll M_{\text{Pl}}$\footnote{This condition is weaker than the one in \cite{Berghaus:2019whh} because thermal friction after the phase transition is suppressed once the quark masses become light, allowing the inflaton to quickly roll to the bottom of its potential.}. 
There is also a bound on $M/g$ from Eq.~\eqref{eq:cond}, since the temperature scales as $M/g$ and the lightest SM quark mass in the UV Higgs vacuum is $m \approx y_u v_\text{UV} \approx 2 \pi y_u T_{\text{RH}}$.
We use this to define the triangular regions in our parameter space plot Fig.~\ref{fig:parameterspace}.
We will see that this upper bound to $M/g$ will overtake the other upper bounds, as since we decrease $T_{\text{RH}}$ we lower the quark masses and $\alpha_S^2 T$ 
is constrained more and more, thus it is this condition that defines the upper limits to our parameter space.

We can individually constrain $M$ and $g$, however, by demanding naturally small quantum corrections to the masses, based on the energy cutoff of the theory $\Lambda_{\text{EFT}} \geq 4 \pi M$. Equivalently, we can let $\Lambda_{\text{EFT}} \equiv 4 \pi \beta M$, where $\beta \geq 1$. These will manifest as \cite{Berghaus:2019whh}

\begin{align}
    \frac{\lambda^2 \Lambda_{\text{EFT}}^2}{16\pi^2} < M^2 &\implies \lambda < \frac{1}{\beta} \, , \label{eq:quantumcorrconstraint1} \\
    \frac{g^2 \Lambda_{\text{EFT}}^2}{16 \pi^2} < m_{\phi}^2 &\implies gM < \frac{m_{\phi}}{\beta} \, . \label{eq:quantumcorrconstraint2}
\end{align}
Combining Eqs.~\eqref{eq:quantumcorrconstraint2} and \eqref{AxionMassMWIHybridInflation} we obtain an upper bound on $g$,
\begin{align}
    g < 2.1 \times 10^{-5} \left(\frac{Q}{10} \right)^{\frac{1}{6}} \left(\frac{\tilde{F}(Q)}{2.4 \times 10^4} \right)^{-\frac{1}{3}} \left(\frac{\mathcal{F}(Q)}{3.5} \right)^{-\frac{3}{4}} \, ,
    \label{gupperbound1}    
\end{align}
which also bounds $\lambda$ from above, 
\begin{equation}
    \lambda \approx 3.3 \times 10^{-2} \left(\frac{Q}{10} \right)^{\frac{1}{3}}  \left(\frac{\tilde{F}(Q)}{2.4 \times 10^4} \right)^{\frac{4}{3}} \left(\frac{\mathcal{F}(Q)}{3.5} \right)^{2} \left( \frac{g}{2.1 \times 10^{-5}} \right)^{4} \left(\frac{M/g}{M_{\text{Pl}}} \right)^2
    \, . \label{SigmaQuarticMWIHybridInflation}
\end{equation}
From Eq.~\eqref{eq:quantumcorrconstraint1}, we also acquire a constraint on $g$ in terms of the upper bound of $M/g$:
\begin{align}
    g < 2.3 \left( \frac{Q}{10} \right)^{-\frac{1}{12}} \left( \frac{\tilde{F}(Q)}{2.4 \times 10^4} \right)^{-\frac{1}{3}} \left( \frac{\mathcal{F}(Q)}{3.5} \right)^{-\frac{1}{2}} \left( \frac{M/g}{M_{\text{Pl}}} \right)^{-\frac{1}{2}} \, . \label{gupperbound2}
\end{align}
Using the constrained theory parameters, we find expressions for the temperature $T$ and Hubble $H$,
\begin{align}
    T &\approx 5.8 \times 10^{-6} \left( \frac{Q}{10} \right)^{1/6} \left( \frac{\tilde{F}(Q)}{2.4 \times 10^4} \right)^{-1/3} \left( \frac{\mathcal{F}(Q)}{3.5} \right)^{-1} \frac{M}{g} \, , \label{TemperatureMWIHybridInflation} \\
    H &\approx 7.0 \times 10^{-10} \left( \frac{Q}{10} \right)^{-1/6} \left( \frac{\tilde{F}(Q)}{2.4 \times 10^4} \right)^{-2/3} \left( \frac{\mathcal{F}(Q)}{3.5} \right)^{-1} \frac{M}{g} \, . \label{HubbleMWIHybridInflation}
\end{align}

Lower bounds on the parameter space arise from remaining within the validity of the EFT which requires $4 \pi f \gtrsim T$, as well as being in the strong regime of warm inflation $Q \gtrsim 10$, which is the regime in which hybrid inflation can match the observed spectral tilt. 
We note that the value of $\alpha_S$ depends on which vacuum the Higgs sits in, and in particular is larger in the UV-vacuum which increases the sphaleron rate and renders more temperatures accessible.
We discuss this subtlety in more detail in Appendix~\ref{app:A}.

We summarize the viable parameter space for hybrid warm inflation with the heavy QCD axion and sphaleron heating in Fig.~\ref{fig:parameterspace}.
The contours in $T_\text{RH}$ indicate the maximum temperature reached after the phase-transition, which is responsible for generating the thermal corrections that allow the Higgs field to relax into the EW-vacuum. 
The benchmark $T_\text{RH}$ values correspond to conservative $1\sigma$ positive variations in the measured top-quark mass, as discussed in Section~\ref{subsec:SMvacua}. 
\edit{Allowing for modified running of the Higgs quartic self coupling from new states populates down to $m_\phi \gtrsim 0.1$ GeV as well -- see Appendix~\ref{subsubsec:lowervacua} for an explicit example.}

It should be emphasized once again that this hybrid setup we have introduced is far from the only possible choice.
Any other choice that achieves the large reheating temperatures necessary to return the Higgs to its EW-vacuum is a potentially viable alternative.
Likewise, even within the context of our model, we chose $g(M/\sqrt{\lambda}) = 0$ in order to mitigate the complications of solving the strong CP problem and to have an observable axion today.
If the coupling between the waterfall field and the axion preserves the solution to the strong CP problem, then this requirement would have been unnecessary, although the resulting axion would be too heavy to observe today.

\section{Constraints and discovery potential}\label{sec:constraints}

Translating current constraints and future projections onto our model space is a nontrivial task, for two reasons. 
First, the region of parameter space that is viable for warm inflation spans axion masses of $10^{-1} \lesssim m_\phi \lesssim 10^5$ GeV, crossing over the scale $\Lambda_\text{QCD}$. 
Our viable region must therefore be split into two regimes, depending on whether QCD is perturbative ($m_\phi \gtrsim 3$ GeV) or whether chiral perturbation theory is valid ($m_\phi \lesssim 1$ GeV), each of which must be treated separately. 
For intermediate masses, $1 \lesssim m_\phi \lesssim 3$ GeV, neither description is accurate, and the validity of any predictions must be treated carefully on a case-by-case basis~\cite{Aloni:2018vki}.
Second, while viable warm inflation sets a value for the coupling to gluons, there is no such restriction on the values of the couplings to electroweak gauge bosons\footnote{
Depending on the details of the UV model, the axion may also couple to any of the SM fermions or the Higgs at tree-level, which would open further search channels -- for example, see~\cite{Bauer:2018uxu,Co:2022bqq}. 
A comprehensive translation of all of these possible constraints is beyond our present scope.
}, $\phi B^{\mu\nu}\widetilde{B}_{\mu\nu}$ and $\phi W_a^{\mu\nu}\widetilde{W}^a_{\mu\nu}$.
The case where either of these couplings exist at tree-level is particularly important in the high mass region where QCD is perturbative.

Let us first consider the low mass region, $m_\phi \lesssim 1$ GeV, where chiral perturbation theory is valid.
After rotating the axion's coupling to gluons into the fermions via a chiral rotation, its couplings to electroweak gauge bosons can be determined by diagonalising the mass matrix of the full $\pi_0-\eta'-\phi$ system, with an additional $\frac{1}{2}m_\phi^2 \phi^2$ term included from the unspecified contribution to the axion potential that makes the axion heavy.
This procedure yields a coupling to photons of~\cite{Alonso-Alvarez:2018irt}
\begin{equation}\label{eq:gagammagammaChiL}
    g_{\phi \gamma\gamma} = -\frac{\alpha_\text{EM}}{2\pi f}\left[\frac{E}{N} - \frac{1}{1-\left(m_\phi/m_{\eta'}\right)^2}\left(\frac{5}{3}+\frac{m_d-m_u}{m_d+m_u}\frac{1}{1-\left(m_\phi/m_\pi\right)^2}\right)\right] \, ,
\end{equation}
where $\alpha_\text{EM}$ is the electromagnetic coupling, $f$ is the axion decay constant, and $E$ and $N$ are the bare axion-photon and axion-gluon couplings, respectively\footnote{That is, the quantized coefficients of the axion couplings to the $U(1)_\text{EM}$ and $SU(3)$ instanton densities.}. 
The masses $m_u$ and $m_d$ are the up and down quark masses, and $m_{\eta'}$ and $m_\pi$ are the masses of the $\eta'$ meson and the $\pi_0$, respectively. 
In writing Eq.~\eqref{eq:gagammagammaChiL}, we have also assumed that $m_\phi$ is overwhelmingly from a source other than QCD dynamics, $m_\phi \gg m_\pi f_\pi / f$. 
This expression has resonances at $m_\phi = m_{\eta'}$ and $m_\phi = m_\pi$, corresponding to points where the mixing is enhanced.
Far below $m_\pi$, these expressions converge back to the usual light axion prediction, while they deviate significantly as one approaches the resonances.
It should be emphasized that since the mixing contribution is $\mathcal{O}(1)$, the cases where $E/N=0$ and where $E/N \neq 0$ are not qualitatively very different, and the resulting constraints in this region are not very model dependent.
The expressions for $g_{\phi WW}$, $g_{\phi ZZ}$, and $g_{\phi Z\gamma}$ may be computed in the same way and are given in~\cite{Alonso-Alvarez:2018irt}, although they turn out to not be relevant for the dominant constraints in our region of interest.

\begin{figure}
    \centering
    \includegraphics[width=0.49\textwidth]{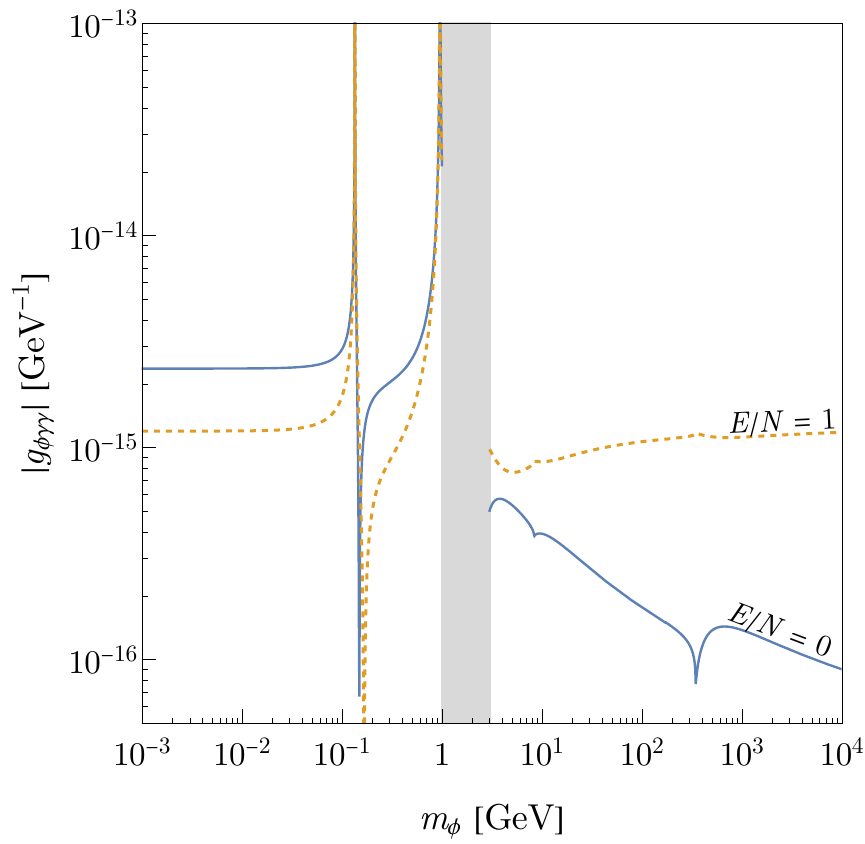}
    \includegraphics[width=0.48\textwidth]{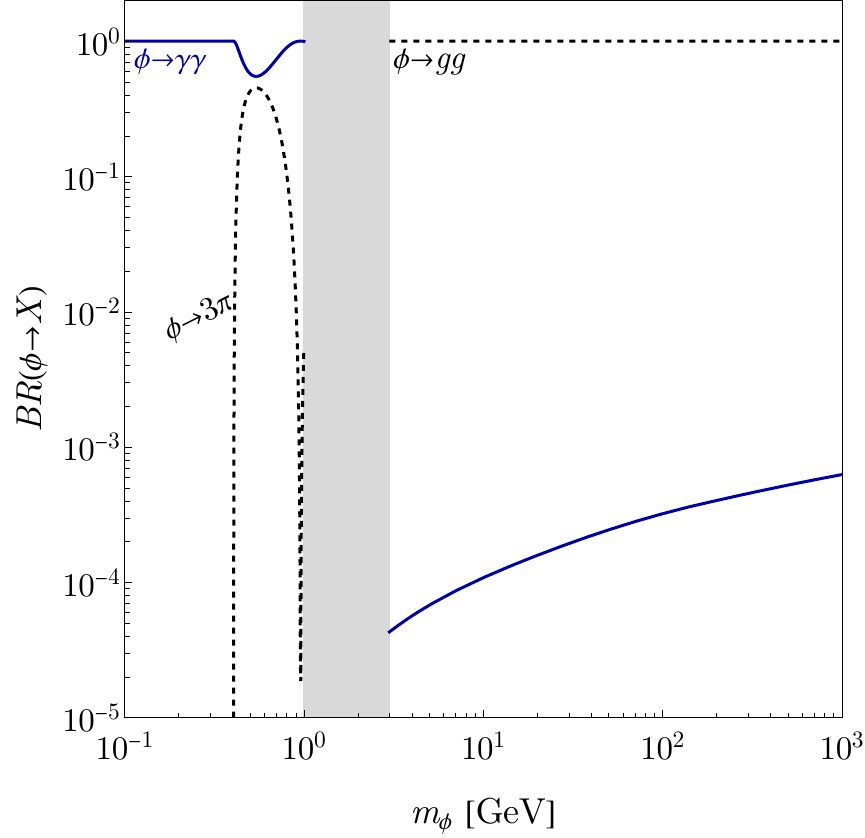}
    \caption{(left) The (approximate) effective axion-photon coupling $|g_{\phi \gamma\gamma}|$ as a function of axion mass for a heavy QCD axion with a representative value $f = 10^{12}$ GeV. We show the cases of no bare coupling to photons $E/N = 0$, as well as $E/N=1$ for comparison. (right) The most important branching ratios as a function of axion mass for a heavy QCD axion with $E/N=1$ and no tree-level couplings to fermions or the Higgs. The gray regions in both plots are where neither the chiral Lagrangian nor perturbation theory admit a valid description of QCD and accurate predictions are difficult. A complete description of this region would be expected to interpolate between the two regimes.}
    \label{fig:gagammagammaConversion}
\end{figure}

At scales $m_\phi \gtrsim 3$ GeV, QCD becomes perturbative, and we may compute the effective couplings to electroweak bosons from loops of fermions and gauge bosons.
At one-loop order, the couplings $g_{\phi VV}$ or $g_{\phi GG}$ generate couplings to fermions which may be probed in flavour experiments, as comprehensively studied in~\cite{Bauer:2021mvw}.
For some constraints, we will need the coupling to photons $g_{\phi \gamma \gamma}$
For an axion that couples only to QCD, this coupling is induced only at two-loop order. 
Explicitly computing the relevant two-loop diagrams is complicated by multiple mass scales, and to our knowledge has not been computed yet.
Nevertheless, we may still estimate the size of the corrections by first considering the one-loop induced coupling to fermions, and then considering this fermion coupling's one-loop contribution to $g_{\phi VV}$~\cite{Bauer:2017ris,Bonilla:2021ufe}.
In particular, doing this yields an estimate for the two-loop contribution to $g_{\phi \gamma\gamma}$ given by~\cite{Bauer:2017ris}
\begin{equation}
    g_{\phi \gamma\gamma} \approx -\frac{\alpha_\text{EM}}{2\pi f}\left[\frac{E}{N}-\frac{3\alpha_S^2}{2\pi^2}\sum_q Q_{q}^2B_1(\tau_q) \log\left(\frac{f^2}{m_q^2}\right)\right] \, ,
\end{equation}
where the sum over $q$ is over the SM quarks, $\tau_q = 4m_q^2/m_\phi^2$, $Q_q$ is the electromagnetic charge of the quark $q$, and $\alpha_\text{EM}$ and $\alpha_S$ are the usual electromagnetic and QCD running couplings, respectively, evaluated at a scale $m_\phi$. The function $B_1(\tau)$ is given by 
\begin{equation}
    B_1(\tau) = 1-\tau f^2(\tau), \qquad f(\tau) =   \Bigg\{  
    \begin{array}{lr}
        \arcsin\frac{1}{\sqrt{\tau}} \ , & \ \tau \geq 1 \ ,   \\
        \frac{\pi}{2} + \frac{i}{2} \ln\frac{1+\sqrt{1-\tau}}{1-\sqrt{1-\tau}} \ , & \ \tau < 1 \ .
    \end{array}
\end{equation}
The contributions to the other couplings $g_{\phi WW}$, $g_{\phi ZZ}$, and $g_{\phi Z\gamma}$ can be estimated in the same way. 
As one may expect, as $m_\phi$ becomes much larger than the meson and light quark masses, the induced effective couplings $g_{\phi VV}$ from mixing effects become suppressed, and so the bare couplings to SM gauge fields are far more important.
Only when the axion has a bare coupling exclusively to QCD are these effects very important, as shown in the left plot of Fig.~\ref{fig:gagammagammaConversion}.

A number of the constraints that we will consider are specific to $\phi \rightarrow \gamma \gamma$ final states. 
However, since our axion couples to QCD, hadronic decay modes are important and suppress $BR(\phi \rightarrow \gamma\gamma)$.
These also reduce the lifetime, which is important for long-lived particle searches and cosmological constraints. 
The decay width for the diphoton final state is given by~\cite{Bauer:2017ris}
\begin{equation}
    \Gamma_{\gamma\gamma} = \frac{g_{\phi \gamma\gamma}^2m_\phi^3}{64\pi}\, ,
\end{equation}
where the difference between the different $m_\phi$ regimes appears in the definition of $g_{\phi \gamma\gamma}$.
For $3m_\pi \lesssim m_\phi \lesssim1$ GeV, the $\phi \rightarrow \pi^a \pi^b \pi^0$ decay channel opens up, with decay rate given by~\cite{Bauer:2017ris}
\begin{align}
    \Gamma_{\pi^a\pi^b\pi^0} &= \frac{1}{6144 \pi^3}\frac{m_\phi}{f^2}\frac{m_\pi^4}{f_\pi^2} \left(\frac{m_d-m_u}{m_d+m_u}\right)^2 g_{ab}\left(\frac{m_\pi^2}{m_\phi^2}\right)\, ,
\end{align}
where $ab = (+-,00)$ and the function $g_{ab}(r)$ is defined as
\begin{align}
\begin{split}
    g_{00}(r) &= \frac{2}{(1-r)^2} \int_{4r}^{(1-\sqrt{r})^2} dz \sqrt{1-\frac{4r}{z}}  \lambda^{1/2}(1,z,r) \, ,\\
    g_{+-}(r) &= \frac{12}{(1-r)^2} \int_{4r}^{(1-\sqrt{r})^2} dz \sqrt{1-\frac{4r}{z}} (z-r)^2 \lambda^{1/2}(1,z,r) \, ,
\end{split}
\end{align}
with $\lambda(x,y,z) = (x-y-z)^2-4yz$.
Finally, at scales $m_\phi \gtrsim 3$ GeV, the primary decay channel is from the decay $\phi \rightarrow gg$, with rate~\cite{Bauer:2017ris}
\begin{equation}
    \Gamma_{gg} =  \frac{\alpha^2_S(m_\phi)m_\phi^3}{32\pi^3 f^2}\left[1+\frac{83}{4}\frac{\alpha_S(m_\phi)}{\pi}\right] \, ,
\end{equation}
which includes the 1-loop QCD correction.  
These are the leading decay modes for our purposes, although as $m_\phi$ approaches 1 GeV some other decay modes that we neglect become important~\cite{Aloni:2018vki}.
For simplicity, we also do not consider possible couplings to fermions and to the Higgs
, which would lead to additional decay channels and further reduce $BR(\phi\rightarrow \gamma\gamma)$ and the lifetime. 
We show the considered branching ratios in the right plot of Fig.~\ref{fig:gagammagammaConversion} for $E/N=1$, where it is clear that for $m_\phi \gtrsim 3$ GeV, any constraints reliant on the decay $\phi \rightarrow \gamma\gamma$ will become significantly suppressed compared to an ALP that does not couple to QCD. 
In the case of $E/N=0$, this would be further exacerbated by the two-loop suppression of $g_{\phi \gamma\gamma}$.

\begin{figure}
    \centering
    \includegraphics[width=\textwidth]{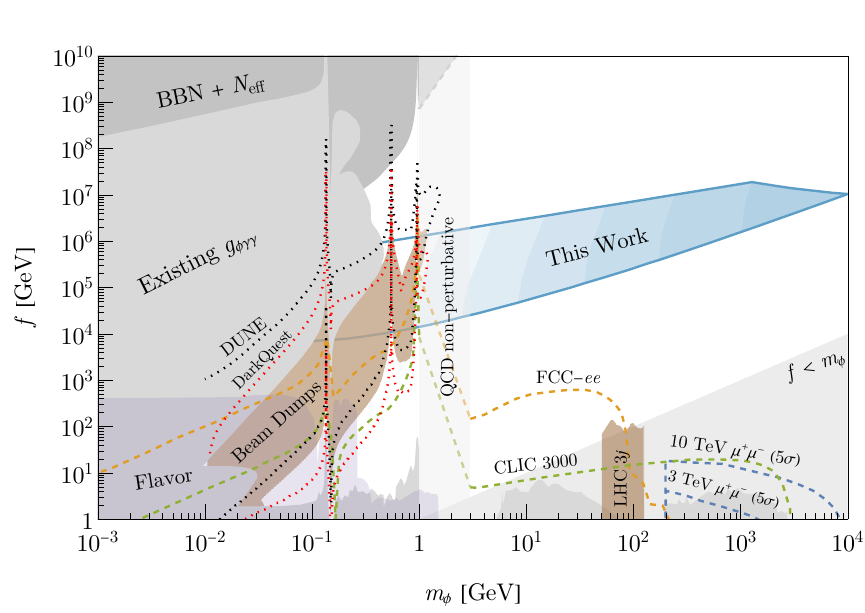}
    \caption{Current and projected axion constraints overlaid onto the parameter space viable for warm inflation, with progressively lighter contours showing conservative $1\sigma$ variations in $m_t$, ignoring any additional uncertainty from the $m_t$ interpretation problem~\cite{Nason:2017cxd,Hoang:2020iah,Workman:2022ynf}. The light gray region between 1 GeV $\lesssim m_\phi \lesssim$ 3 GeV is where QCD is non-perturbative and perturbation theory cannot make accurate predictions, while the light gray region covering the bottom right corner shows where the axion EFT breaks down. The gray regions are the combination of current $g_{\phi \gamma\gamma}$ constraints~\cite{AxionLimits} from astrophysics~\cite{Lucente:2020whw,Caputo:2022mah,Hoof:2022xbe,Muller:2023vjm,Dev:2023hax,Diamond:2023cto} and collider searches~\cite{Jaeckel:2015jla,Knapen:2016moh,Dolan:2017osp,CMS:2018erd,Aloni:2019ruo,ATLAS:2020hii,Belle-II:2020jti,BESIII:2022rzz}, while the darker gray region is from cosmology~\cite{Depta:2020wmr}. The brown region below $1$ GeV is a combination of several beam dump experiments~\cite{Blumlein:1990ay,BERGSMA1985458,E137,E141}. The purple region shows loop-induced flavour constraints~\cite{Bauer:2021mvw}. The small brown region near $100$ GeV shows the current $g_{\phi GG}$ LHC bound from 3-jet events~\cite{Mariotti:2017vtv}. Projected constraints from DUNE~\cite{Brdar:2020dpr} and DarkQuest~\cite{SeaQuest:2017kjt,Berlin:2018pwi,Batell:2020vqn} are overlaid as dotted lines. Projections for FCC-$ee$~\cite{Bauer:2018uxu}, CLIC-3000~\cite{Bauer:2018uxu}, and a $\mu^+\mu^-$ collider~\cite{Bao:2022onq,Han:2022mzp} on $\phi\rightarrow \gamma\gamma$ processes assuming $E/N=1$ are shown as dashed lines. All other constraints assume $E/N=0$, i.e. the axion couples only to QCD at tree-level.}
    \label{fig:Constraints}
\end{figure}

We are finally in a position to present current constraints and future projections on our axion parameter space.
We show these constraints overlaid on the parameter space in which the heavy QCD axion can function as the warm inflaton in Fig.~\ref{fig:Constraints}. 
For most of these constraints, we assume the axion couples only to QCD at tree-level\footnote{Considering i.e. $E/N=1$ instead would only cause a small numerical change in these constraints without changing any qualitative conclusions.}, $E/N=0$.
The $\mu^+\mu^-$ and $e^+e^-$ collider constraints are the sole exceptions, since there is a rather large difference between $E/N=0$ and $E/N=1$.
For these constraints we assume $E/N = 1$ to demonstrate the more optimistic reach.
The blue region labelled `This work' corresponds to the viable parameter space of our model from Fig.~\ref{fig:parameterspace}, where the heavy QCD axion is the inflaton of warm inflation, with the progressively lighter blue bands indicating positive $1\sigma$ shifts of $0.3$ GeV in $m_t$, conservatively neglecting any additional uncertainty from the $m_t$ interpretation problem~\cite{Nason:2017cxd,Hoang:2020iah,Workman:2022ynf}. 
Additional BSM contributions that would generate a lower $v_\text{UV}$ than in the SM, such as the VLF example discussed in Section~\ref{subsubsec:lowervacua}, would also populate the lighter blue parameter space.
The light gray region in the bottom right indicates the breakdown of our EFT\footnote{This does not mean that this parameter space is excluded, rather it means that the full UV model needs to be considered. For example, in a Peccei-Quinn scenario, one would need to include the radial mode and the additional heavy fermion(s) as dynamical degrees of freedom.}, $f < m_\phi$.
In a lighter gray, we also show the region 1 GeV $<m_\phi <$ 3 GeV where nonperturbative QCD effects become dominant. 
This region is not excluded; rather, accurately mapping from a constraint on a coupling $g_{\phi VV}$ to the scale $f$ is nontrivial and must be handled with care~\cite{Aloni:2018vki}, and some constraints may be significantly altered there. 
Current experimental constraints are shown as filled gray, purple, and brown contours, while experimental projections are shown as dashed lines. 
These come from a number of different astrophysical, cosmological, and collider sources, which we will discuss in turn.

The gray region labelled `Existing $g_{\phi \gamma\gamma}$' consists of current constraints on the coupling to photons, as collected at~\cite{AxionLimits}. 
We show only the most dominant constraints to simplify the presentation. 
For low $f$ (high $g_{\phi \gamma\gamma}$), these primarily consist of various collider searches~\cite{Jaeckel:2015jla,Knapen:2016moh,Dolan:2017osp,CMS:2018erd,Aloni:2019ruo,ATLAS:2020hii,Belle-II:2020jti,BESIII:2022rzz}, which are unfortunately several orders of magnitude too weak to probe our region of interest. 
For $m_\phi \lesssim 1$ GeV and moderate $f$, the constraints are dominated by  supernovae~\cite{Lucente:2020whw,Caputo:2022mah,Hoof:2022xbe,Muller:2023vjm} (see also~\cite{Caputo:2021rux, Caputo:2022rca, Diamond:2023scc} for some recent updates) and GW170817~\cite{Dev:2023hax,Diamond:2023cto}.
For these we have assumed that the important physics is captured primarily by the $g_{\phi \gamma\gamma}$ induced by mixing with the pion. 
This is clearly appropriate below the pion mass, however between $3m_\pi \lesssim m_\phi \lesssim 1$ GeV, these stellar constraints may be altered by an $\mathcal{O}(1)$ factor due to the possible decay $\phi \rightarrow 3\pi$ that we neglected.

In dark gray, we show the cosmological bounds from BBN and $N_\text{eff}$~\cite{Depta:2020wmr}, which are dominant for $f\gtrsim 10^8$ GeV. 
For this constraint, we have used the most aggressive bound from~\cite{Depta:2020wmr}, and taken into account the change in the axion lifetime from additional hadronic decays. 
Especially above 3 GeV, these decay modes dominate the lifetime and weaken the constraint significantly compared to the naive expectation when including only $\phi \rightarrow \gamma\gamma$.
The dashed line from $1\lesssim m_\phi \lesssim 3$ GeV is to emphasize that the constraint in that region would be significantly altered due to the presence of additional hadronic decay modes that we have neglected, although it should interpolate between the two regimes.

In brown and purple, we show the constraints arising from the coupling $g_{\phi GG}$ itself. 
In purple, we show the flavour constraints as presented in~\cite{Bauer:2021mvw}. 
The presence of a coupling to gluons generates a coupling to fermions from renormalization-group running from the scale $f$ to $m_\phi$. 
While this is loop suppressed, it generates flavour changing and exotic decay signatures, which are very well constrained, although our parameter space remains out of reach.
The brown region at high masses, $m_\phi \approx 100$ GeV is from $3j$ searches at the LHC~\cite{Mariotti:2017vtv}. 
The brown region at lower masses is the combination of the beamdump experiments NuCal~\cite{Blumlein:1990ay}, CHARM~\cite{BERGSMA1985458}, E137~\cite{E137}, and E141~\cite{E141}, which we obtained using ALPINIST~\cite{Jerhot:2022chi}.
As dotted lines, we also show projections for DUNE~\cite{DUNE:2021tad,Kelly:2020dda} and DarkQuest~\cite{SeaQuest:2017kjt,Berlin:2018pwi,Batell:2020vqn}, again using ALPINIST. 
The constraints from SHIP~\cite{SHiP:2015vad} and SHADOWS~\cite{Baldini:2021hfw} would cover a very similar region; we have neglected them for ease of presentation.
These experiments will begin to probe the low $m_\phi$ parameter space relevant for our model.

Finally, we include projections for the reach at future lepton colliders, assuming $E/N=1$, since the $E/N=0$ case would have no tree-level production mechanism and therefore limited constraining power.
These consist of FCC-$ee$~\cite{Bauer:2018uxu}, CLIC-3000~\cite{Bauer:2018uxu}, and a $\mu^+\mu^-$ collider~\cite{Bao:2022onq,Han:2022mzp}.
The $\mu^+\mu^-$ collider results are $5\sigma$ discovery limits, as no 2$\sigma$ constraint projections in the $(C_{\phi VV},m_\phi)$ plane were presented in~\cite{Bao:2022onq,Han:2022mzp}.
All of these constraints have taken into account the altered $BR(\phi \rightarrow \gamma \gamma)$ from hadronic decay modes as shown in Fig.~\ref{fig:gagammagammaConversion}.
Since hadronic decays are dominant, searches for resonances in hadronic final states can likely push the constraining power considerably further than the plot suggests, potentially probing our parameter space. 
However, such projections unfortunately have not been computed yet to our knowledge. 
We leave this to future work. 

Beyond collider signatures of the heavy QCD axion, a cosmological avenue towards the discovery of warm inflation lies in its imprint on non-gaussianities. In particular \cite{Mirbabayi:2022cbt}
identifies a new template for the warm bispectral shape that dominates in the strong regime of warm inflation, and increases with higher $Q$. Their template has a $60\%$ correlation with one of the bispectral shapes of warm inflation proposed in \cite{Bastero-Gil:2014raa}, but predicts a larger amplitude of non-gaussianities. It would be interesting to derive constraints from Planck \cite{Planck:2019kim} on the new template calculated in \cite{Mirbabayi:2022cbt}, and forecast the sensitivity with EUCLID \cite{Amendola:2016saw,  refId0}, SPHEREx \cite{SPHEREx:2018xfm}, or HI intensity mapping experiments such as SKA-I \cite{SKA:2018ckk, Camera:2014bwa, Li:2017jnt}. Based on the new prediction in \cite{Mirbabayi:2022cbt}, these missions may be able to rule out or make a discovery in the region above $Q > 100$.

\section{Conclusions}
\label{sec:concl}
In this work we have proposed a heavy QCD axion as the inflaton of warm inflation, where its coupling to QCD sources the radiation bath via sphaleron heating, maintaining finite temperature during inflation. The radiation bath is made up of SM gluons in their non-confined phase. Light quarks which suppress sphaleron heating are absent in the early universe in our model due to the Higgs field residing in a UV-vacuum during warm inflation. The mechanism presented in this paper to make SM quarks heavy during warm inflation relies on a large reheating temperature after warm inflation, $T_\text{RH} > v_\text{UV}$. We propose hybrid warm inflation to provide the large reheating temperature which allows the Higgs field to relax to its EW-vacuum. Hybrid inflation also suppresses the slow-roll parameter $\epsilon_V$, enabling warm inflation with sphaleron heating to reproduce the observed spectral tilt without tuning. 

We identify a region in the heavy axion $f - m_{\phi}$ plane in which our model is viable, and consolidate existing and future reach in the heavy axion parameter space from astrophysical, cosmological, and collider sources, finding that future beam dump experiments such as DarkQuest are able to probe the lower mass region of our model. Including hadronic final states in the forecast of future collider capabilities may significantly enhance the reach, potentially probing the intermediate mass region.
On the cosmological side, non-gaussianities with a bispectral shape predicted by warm inflation could be a smoking gun for the discovery of warm inflation that may be within reach with upcoming experiments. 

In this work we have presented a proof-of-principle model in which a heavy QCD axion and its coupling to gluons can maintain warm inflation. 
There may be additional model realizations that lead to a similar phenomenon. 
\edit{Our work provides a roadmap for this class of models by 
identifying and overcoming the main model building challenges towards embedding warm inflation in the SM. The challenges are light quarks in the SM which suppress sphaleron heating, as well as the necessity of a UV-potential in order for the QCD axion to play the role of the inflaton. }

\edit{Specifically, our example requires a large reheating temperature after warm inflation, which we achieve via a hybrid warm inflation setup. This has been previously shown to be a compelling model in which warm inflation can accommodate the measurements of the spectral tilt. }
\edit{Our model shows that warm inflation can be viable by producing the thermal bath from coupling directly to Standard model particle content. On the downside, our concrete example does not allow for a smooth exit into a SM radiation dominated era. Hybrid inflation also already features a suppression of the tensor-to-scalar ratio, which is one of the interesting consequences of warm inflation. Notably however, cold hybrid inflation is not compatible with observations due to its predicted blue tilt, thus making hybrid inflation warm is one avenue to restore its concordance with measurements.} 

\edit{Lastly, we have connected the UV-potential necessary for the QCD axion to be the inflaton with proposed UV completions that preserve it as a solution to the strong CP problem. We point out that the presence of such a UV-potential also addresses the axion quality problem. }



\edit{Twenty five years ago it was questioned whether warm inflation is possible \cite{Yokoyama:1998ju}. Several proposals have since been able to overcome the challenge to achieve efficient particle production without a large backreaction on the inflaton potential.
 Here we have shown that coupling an inflaton directly to the SM particle content can already lead to warm inflation in the presence of few BSM ingredients, taking a step towards fully embedding warm inflation in the SM. 
}

\acknowledgments
We are grateful to Hooman Davoudiasl, Marco Drewes, Peter Graham, Junwu Huang, Saarik Kalia, Seth Koren, Patrick Meade, Ryan Plestid, Marvin Schnubel, Robert Szafron, Mark Wise, and Sebastian Zell for helpful discussions. We further thank Saarik Kalia, Junwu Huang, and Anson Hook for comments on the manuscript.  
K.B. acknowledges the support of NSF Award PHY2210533, and thanks the U.S. Department of Energy, Office of Science, Office of High Energy Physics, under Award Number DE-SC0011632 and the Walter Burke Institute for Theoretical Physics. 
The work of M.F. was supported in part by the National Science Foundation grant PHY-2210533. 
M.F. is also supported by the U.S. Department of Energy, Office of Science, Office of Workforce Development for Teachers and Scientists, Office of Science Graduate Student Research (SCGSR) program. The SCGSR program is administered by the Oak Ridge Institute for Science and Education (ORISE) for the DOE. ORISE is managed by ORAU under contract number DE-SC0014664.

\appendix

\section{\textcolor{black}{Viable Higgs vacua with $v_{\text{UV}} \lesssim 10^9 \, \text{GeV}$}}\label{subsubsec:lowervacua}

When using the SM running of $\lambda_H$, conservatively varying $m_t$ by up to $1.5$ GeV cannot yield a $v_\text{UV}$ smaller than $\sim 10^9$ GeV.
However, parameter space corresponding to lower $v_\text{UV}$ and $T_\text{RH}$ is easier to access with current and near future axion searches, as \edit{shown} in Sec.~\ref{sec:constraints}.
One way for a second minimum to appear sooner is if the running of $\lambda_H$ is altered to become negative sooner than in the SM. 
A simple realization is the addition of vector-like fermions (VLFs) that couple to the Higgs through Yukawa interactions. 
Just like in the case of the top-quark, these give the correct sign to $\lambda_H$'s $\beta$-function to make it negative faster.

To be concrete, let us consider a simple singlet-doublet (SD) model. Similar to~\cite{Cheng:2019qbd}, we consider two singlet Weyl fermions $S_{L,R}$ with hypercharge $Y=0$ and two $SU(2)_L$ doublets $D_{L,R}$ with hypercharge $Y=1/2$. 
Ignoring any mixing effects with the SM fermions\footnote{This can always be done by imposing an additional global symmetry under which only the new fermions are charged.}, the relevant interaction terms are
\begin{equation}
 \mathcal{L}_\textrm{SD} \supset -M_D \bar{D}_L D_R - M_S \bar{S}_L S_R -y_1 \bar{D}_L H S_R -y_2 \bar{D}_R H S_L  + h.c.
\end{equation}
Singlet-doublet models have been looked at in the literature in a variety of contexts including as a dark matter candidate~\cite{Mahbubani:2005pt,DEramo:2007anh,Enberg:2007rp,Cohen:2011ec,Cheung:2013dua,Abe:2014gua,Calibbi:2015nha,Freitas:2015hsa,Banerjee:2016hsk,Cai:2016sjz,LopezHonorez:2017zrd,Fraser:2020dpy,Asadi:2022xiy,Wang:2022dte}, though for our purposes only its effects on the Higgs potential are relevant.
These effects have been examined in~\cite{Wang:2018lhk} for a Majorana singlet and in~\cite{Cheng:2019qbd} for the Dirac singlet we consider here. 
The contributions of $y_1$ and $y_2$ to $\beta_{\lambda_H}$ are given by
\begin{align}
\begin{split}
    \delta\beta_{\lambda_H}^{(1)} = \frac{1}{(4\pi)^2}\bigg[&4\lambda_H\left(y_1^2+y_2^2\right)-2y_1^4-2y_2^4\bigg] \, ,\\
    \delta\beta_{\lambda_H}^{(2)} = \frac{1}{(4\pi)^4}\bigg[&\left(-\frac{g_1^4}{4}-\frac{g_1^2 g_2^2}{2}+\frac{5 g_1^2 \lambda_H }{2}-\frac{3 g_2^4}{4}+\frac{15 g_2^2 \lambda_H }{2}-48 \lambda_H^2\right)\left(y_1^2 + y_2^2\right) \\
    & - \lambda_H\left(y_1^4 + y_2^4\right) + 10\left( y_1^6 + y_2^6\right)\bigg] \, ,
\end{split}
\end{align}
where the superscript is the loop order and $(g_1,g_2)$ are the $U(1)_Y$ and $SU(2)_L$ gauge couplings, respectively.
As anticipated, the dominant contributions of $y_1$ and $y_2$ come along with a negative sign, making $\lambda_H$ run negative faster.
We will consider $y \equiv y_1 = y_2$ and $M\equiv M_S = M_D$ for simplicity, although the results would be very similar for more general values of $y_1$ and $y_2$ since they contribute identically to $\beta_{\lambda_H}$. 
Likewise, relaxing the mass equality would not meaningfully change our results on scales $\mu \gg M_S,M_D$ unless there is a large hierarchy between the two masses.
The physical masses in this limit are given by
\begin{equation}
    M^2_{1,2}(h) = M^2 + \frac{y^2}{2}h^2 \pm 2\sqrt{2} y M h \, .
\end{equation}

We use SM $\beta$-functions up to the scale $M_1\approx M_2 \approx M$.
Once past the threshold $M$, we include the contributions to the $\beta$-functions from the new vector-like fermions, and we likewise include the usual one-loop contributions to the effective potential
\begin{equation}
    \lambda_\text{eff}^{\textrm{SD}}(h) = \lambda_\text{eff}^\text{SM}(h) + \sum_i \frac{1}{64\pi^2}y^4\left[\log{\frac{M^2_{i}(h)}{\mu^2}}-\frac{3}{2}\right] \, ,
\end{equation}
where we have once again absorbed these new terms into an effective quartic coupling, neglecting the terms quadratic in $h$ that are irrelevant at scales $h \gg M$.
We perform the matching at the threshold $M$ using the same procedure as~\cite{Blum:2015rpa} by expanding the above contribution into powers of $h^n$, and requiring that the potential is continuous at $M$. 
This yields finite changes in $\lambda_H$ and $m_H^2$ at the matching threshold, as well as an effective $h^6$ interaction when at scales $\mu < M$ that is numerically unimportant.

\begin{figure}
    \centering
    \includegraphics[width=0.8\textwidth]{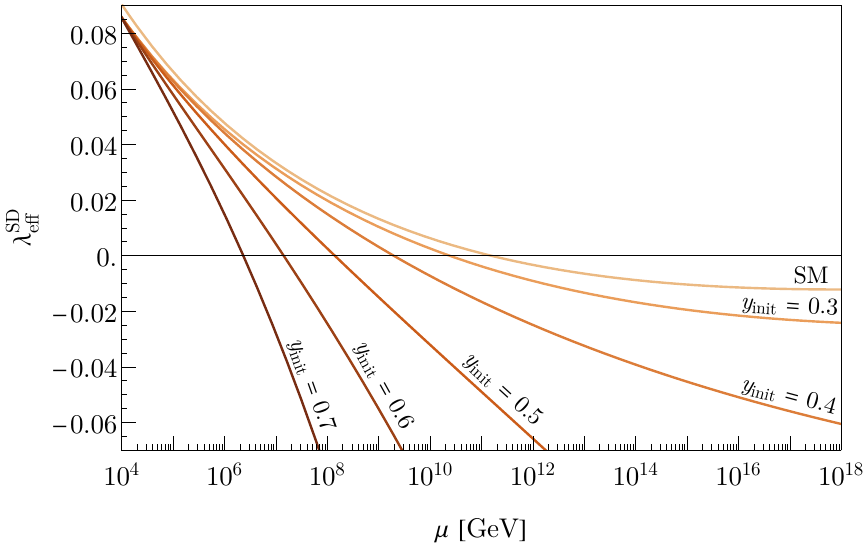}
    \caption{The running of the effective quartic coupling $\lambda_\text{eff}^\text{SD}$ in the singlet-doublet model discussed in the text for different initial values $y_\text{init}$ of the new heavy fermion Yukawa coupling with $M=10$ TeV. For small $y_\text{init}$, the running of $\lambda_\text{eff}^\text{SD}$ and corresponding $v_\text{UV}$ range is nearly the same as in the SM while for larger initial values $v_\text{UV}$ may be several orders of magnitude smaller.}
    \label{fig:SDrunning}
\end{figure}

In order to ensure the stability of our vacuum, in principle one should explicitly compute the bounce action and corresponding tunneling rate for every potential configuration. 
This is particularly important when determining what ranges of $y$ are stable, metastable, or unstable, as in~\cite{Wang:2018lhk,Cheng:2019qbd,Gopalakrishna:2018uxn}.
However, in our case, we introduce a new operator to stabilise the potential after it becomes negative. 
This generally makes unstable configurations much more long-lived, as was explicitly shown in~\cite{Gopalakrishna:2018uxn}. 
The precise value at which this operator must dominate is not particularly important, so long as there is enough room between the scale when $\lambda_\textrm{eff}^\textrm{SD}= 0$ and when the new operator must turn on so that it is not terribly fine-tuned. 
We will therefore use the approximation that the scale at which the new operator saves the potential is given by the scale at which $\lambda_\textrm{eff}^\textrm{SD} = -0.07$, roughly corresponding to the onset of vacuum instability~\cite{Blum:2015rpa}. 
A more precise calculation of the tunneling rate may change this position somewhat, but it would not change our overall conclusions.

The impact of the new fermions on the running of $\lambda_\text{eff}^\text{SD}$ are exemplified in Fig.~\ref{fig:SDrunning} for various initial values of $y_\text{init} \equiv y(\mu=M)$ and a benchmark mass of $M=10$ TeV.
We see that for perturbatively small initial values $y_\text{init}$, it is possible to make $\lambda_\text{eff}^\text{SD}$ become negative several orders of magnitude sooner, enough to populate the entire parameter space we present in Sec.~\ref{sec:warmhybridinflation}.
For smaller values of $M\sim 1$ TeV, the scale $\mu$ at which $\lambda_\text{eff}^\text{SD}$ crosses zero lowers by roughly another order of magnitude.
Likewise, there is over an order of magnitude between when $\lambda_\text{eff}^\text{SD}=0$ and $\lambda_\text{eff}^\text{SD} \approx -0.07$ so long as $y_\text{init} \lesssim 0.8$, meaning that the dimension six operator $(H^\dagger H)^3$ has an order of magnitude to become important and is not fine tuned.

We should emphasize that the behaviour we discuss is generic for fermionic extensions, since Yukawa couplings always contribute with the correct sign to induce this running.
The quantum numbers of the fermions are also not particularly important, so long as the mass is heavy enough to evade collider searches and they do not form exotic stable relics after reheating. 
UV completions of axion models often invoke new heavy fermions, and similarly the axion often arises from a gauge singlet complex scalar, from which the dimension six operator could arise after integrating out the radial mode. 
It would be particularly interesting if a viable Peccei-Quinn UV completion could be developed to realise the axion, fermions, and dimension six operator simultaneously. 
Achieving this is complicated by the fact that our axion must be heavy, which requires rather careful modelbuilding, as we review in Sec.~\ref{sec:heavyQCDaxions}.
We leave this interesting direction for future work.

\section{The QCD coupling in the UV Higgs vacuum}\label{app:A}
During inflation, sphaleron heating via the thermal friction coefficient $\Upsilon_\text{sph} = \kappa \alpha^5 T^3/f^2$ depends crucially on the value of the coupling $\alpha$. 
Naively, since these are SM QCD sphalerons, we could simply insert the known running of $\alpha_S$ to determine the rate.
However, the value of $\alpha_S$ at scales below $v_\text{UV}$ is altered by the quarks being heavy during inflation. 
To see how it changes, consider the following. 
In the deep UV where all particles are approximately massless, the differences between the two vacua should disappear. 
At energy scales below $v_\text{UV}$, there are two distinct IR effective theories depending on which vacuum the Higgs sits in, both of which emerge from the same UV theory in a calculable way.
The correct approach is then to first determine the couplings in the UV from measurements in our EW vacuum, then match those couplings onto the different IR theory where the Higgs lives in the new minimum.

In our results, we therefore first take the measured values of all SM couplings at the scale $m_t$ and evolve them up to a scale $\mu \gg v_\text{UV}$ using the full RGE as discussed in Section~\ref{subsec:SMvacua}.
We then flow back to scales $\mu \ll v_\text{UV}$, where the altered quark masses from the UV minimum lead to different matching scales for $\alpha_S$ when they are integrated out.
When running back, we do this at 1-loop for simplicity since $\alpha_S$ is perturbatively small at high scales, and at 1-loop we only need to consider the matching conditions for $\alpha_S$ rather than the full set of SM couplings. 
The matching thresholds occur sooner in $v_\text{UV}$ than in $v_\text{EW}$, and since the running of $\alpha_S$ with $n_f = 6$ flavours is slower than for $n_f < 6$ flavours, the value of $\alpha_S$ at scales $\mu < v_\text{UV}$ is \textit{larger} in the UV-vacuum than in the electroweak vacuum, enhancing the sphaleron rate.

\begin{figure}
    \centering
    \includegraphics[width=0.495\textwidth]{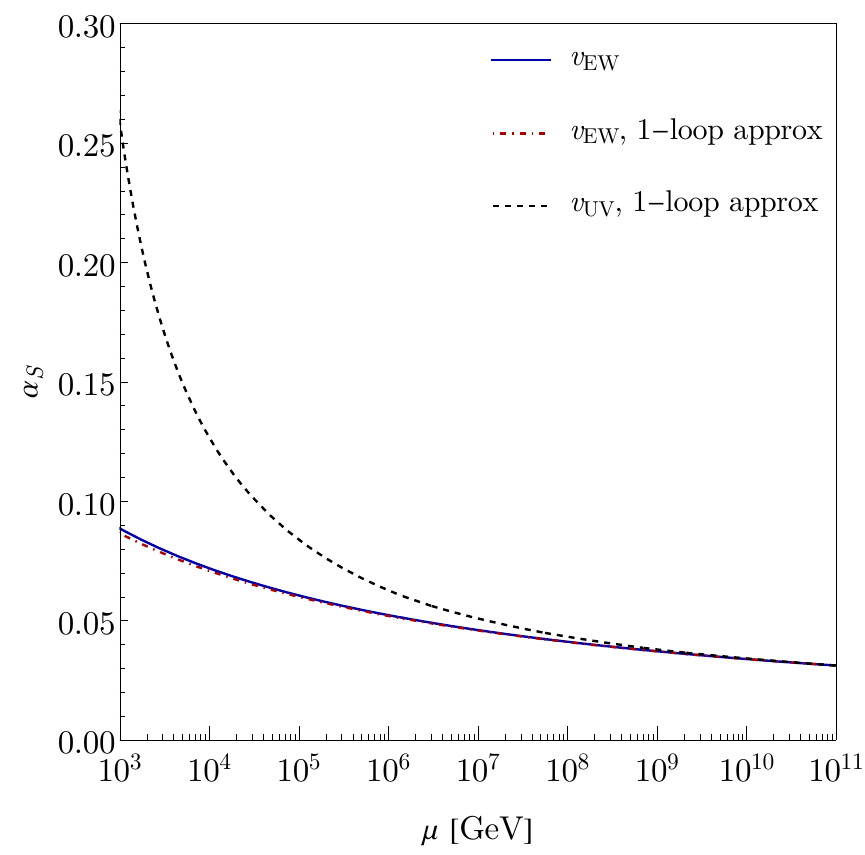}
    \includegraphics[width=0.49\textwidth]{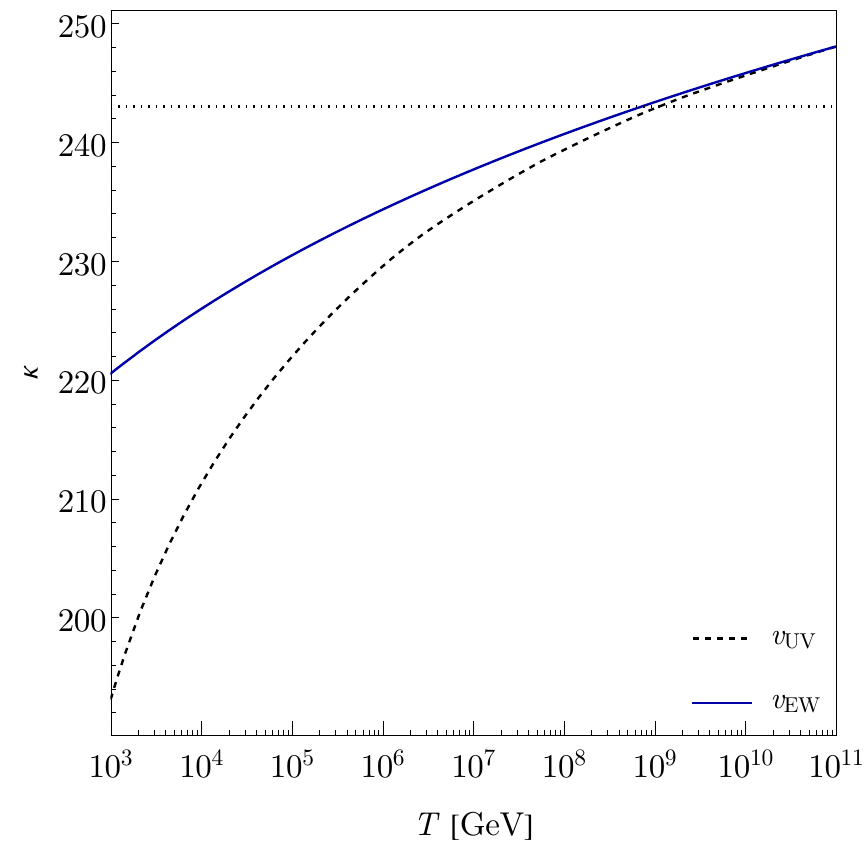}
    \caption{(\textit{left}) The value of the QCD coupling $\alpha_S$ as a function of scale in the different Higgs phases. The blue curve shows the SM value in our EW vacuum $v_\text{EW}$, while the black shows the change due to larger quark masses when in the smallest possible UV-vacuum $v_\text{UV}$ for $m_t = 172.69$. In the UV-vacuum, running down from the UV to the IR is done at 1-loop order. For comparison, we show this approximation in red for the EW vacuum, where it makes little numerical difference. (\textit{right}) The running with temperature of the dimensionless proportionality constant $\kappa$ entering into the sphaleron rate in Eq.~\eqref{eq:sphaleronrate}, in the EW and UV vacua. The dotted line indicate the estimate $\kappa = N_c^5$.}
    \label{fig:alphas}
\end{figure}

We show the value of $\alpha_S$ in the two minima in Fig.~\ref{fig:alphas}, where the enhancement over the calculation in the EW minimum is clearly significant.
Since we only compute the RGE to 1-loop in the UV minimum when flowing down, we show this same approximation for the EW minimum as well, where it agrees to at least the $\sim 1\%$ level for all shown scales. 
Of course, the corrections in the UV minimum would be more important since $\alpha_S$ is larger, but we take this to be a reasonable approximation. 
The running of $\alpha_S$ depends on the value of $v_\text{UV}$, which we evaluate at $v_{\text{UV}} = $ for the black curve shown in Fig.~\ref{fig:alphas}. We use the black curve for the evaluation of the parameter space in Fig.~\ref{fig:parameterspace}, and neglect the small deviations due to variation in $v_{\text{UV}}$.

Having determined the running of $\alpha_S$ with energy scale, we use it to calculate the sphaleron rate's proportionality constant $\kappa$ using the prescription in \cite{Moore:2010jd}. In Sec.~\ref{sec:warmhybridinflation}, a factor of $\sqrt{\kappa \alpha_S^5}$ appears in the final expression for the axion decay constant $f$ as shown in Eq.~\eqref{ACDMWIHybridInflation}. During inflation the relevant energy scale is the temperature $T$, and so, as we sit in the UV Higgs vacuum, we run $\alpha_S$ as depicted in Fig.~\ref{fig:alphas}. Then from \cite{Moore:2010jd}, we calculate $\kappa$ for SU($N_c = 3$) with  $\alpha_S = g^2/(4 \pi)$ as 

\begin{align}
    \kappa = (132 \pm 4) \left( \frac{g^2 T^2}{m_{\text{D}}^2} \right) \left( \ln \left( \frac{m_{\text{D}}}{\gamma} \right) + 3.041 \right),
\end{align}

\noindent where $\gamma$ is found by numerically solving

\begin{align}
    \gamma = \frac{N_c g^2 T}{4 \pi} \left( \ln \left( \frac{m_{\text{D}}}{\gamma} \right) + 3.041 \right), \\
    m_{\text{D}}^2 = \frac{2N_c + N_f}{6}g^2 T^2.
\end{align}
Solving these equations, we find the result for $\kappa$ shown in Fig.~\ref{fig:alphas}, where we set $N_c = 3$ and $N_f = 0$ since the quark masses are heavier than $T$ and do not contribute to the sphaleron dynamics. We see that in the UV-vacuum $\kappa$ is slightly lowered at smaller temperatures, while the combination $\kappa \alpha_S^5$ is overall increased since $\alpha_S$ is enhanced, effectively increasing the parameter space that is compatible with the EFT bound $T \lesssim 4 \pi f$.

\section{UV completion of the warm hybrid inflation potential}
\label{sec:UV}
The warm hybrid inflation potential we employ as concrete realization of warm inflation with the heavy QCD axion
\begin{align}
\label{eq:B1}
    V(\sigma,\phi) = \frac{1}{4\lambda}(M^2 - \lambda |\sigma|^2)^2 + \frac{1}{2}g(\sigma)^2 |\sigma|^2 \phi^2 + \frac{1}{2}m_{\phi}^2 \phi^2 \, ,
\end{align}
relies on a waterfall field dependence in the coupling $g(\sigma)$, such that $g(0) = g$, and $g(\sim M/\sqrt{\lambda}) =0$.
As a concrete UV completion we consider a dark $SU(N_c^d)$ gauge group with two vectorlike quarks, $\tilde{u} +\tilde{u}^c$ and $\tilde{d} + \tilde{d}^c$ charged under $U(1)_{\sigma}$ which was proposed in \cite{Gong:2021zem}. The interactions are given by
\begin{equation}
\mathcal{L} \supset m_{\tilde{u}} \tilde{u} \tilde{u}^c + y \sigma \tilde{u}^c d + y' \sigma^{*} \tilde{u} \tilde{d}^c + m_{\tilde{d}} \tilde{d} \tilde{d}^c +\frac{\alpha_d \phi }{16 \pi^2 f_{d}} G^d_{\mu \nu} \tilde{G}^{d \mu \nu} + \frac{1}{2} m_\phi^2 \phi^2,  
\end{equation}
where $m_\phi^2$ is the heavy mass contribution that is assumed to solve the strong CP problem as discussed in Sec.~\ref{sec:heavyQCDaxions}, and the dark confining scale is $m_{\tilde{d}} \ll \Lambda_d \ll m_{\tilde{u}}$.
Integrating out the heavy quark one obtains a $\sigma$-dependent effective quark mass for the lighter quark 
\begin{equation}
\left(\frac{y y'}{ m_{\tilde{u}}} |\sigma|^2 + m_{\tilde{d}} +\delta{m_{\tilde{d}}} \right) \tilde{d} \tilde{d}^c \, ,
\end{equation}
where $\delta m_{\tilde{d}} = \frac{y y'}{16 \pi^2} m_{\tilde{u}} \log(\Lambda_{\text{EFT}}^2/M^2)$ accounts for the radiative corrections to the lighter dark quark, and $\Lambda_{\text{EFT}}$ is the cutoff of the theory. For $\Lambda_{\text{EFT}} \geq 4 \pi M$, yukawa couplings of order $y \sim y' \leq 0.6$ maintain the seperation of scales.

During warm inflation, while $\langle \sigma \rangle = 0$, the mass of $\tilde{d} + \tilde{d}^c$ is lighter than the confinement scale $\Lambda_d$, thus it confines to form a meson $\langle \tilde{d} \tilde{d}^c \rangle$ with a mass and decay constant $\sim \Lambda_d$. The axion inflaton mixes with the dark meson due to its anomalous coupling to the dark gauge bosons. Integrating out the dark meson, one obtains the term 
\begin{align}
 V(\phi,\sigma) & \supset  -\left| \frac{y y'}{m_{\tilde{u}}} \right| \Lambda^3_d \cos\left({\frac{\phi}{f_d} + \beta}\right)  |\sigma|^2  \nonumber \\
&  \supset -\frac{1}{2} \delta M^2 |\sigma|^2  + g^2 \phi^2 |\sigma|^2 \, 
\end{align}
where $\delta M^2 = 2 \left| \frac{y y'}{m_{\tilde{u}}} \right| \Lambda^3_d$, and $g^2 =  \left| \frac{y y'}{m_{\tilde{u}}} \right| \frac{\Lambda^3_d}{2f_d^2}$. 
After inflation when the waterfall field obtains a large expectation value $\langle \sigma \rangle = M/\sqrt{\lambda} $, the lighter quark's mass, $m^{\text{eff}}_{\tilde{d}} \approx yy'/m_{\tilde{\mu}} \langle\sigma \rangle ^2 \gg \Lambda_d $, increases beyond the confinement scale $\Lambda_d$. 
In the absence of mesons the mixing with the axion vanishes and we have exactly $g^2 = 0$ \cite{Gong:2021zem}. 
Integrating out the heavy quarks then leaves a pure Yang-Mills with a contribution to the axion mass of $\delta m^2_{\phi} \approx \tilde{\Lambda}^4_d/f^2_d$, where $ \Lambda_d \lesssim \tilde{\Lambda}_d$ is the confinement scale after the waterfall transition, which is generally larger than $\Lambda_d$ by the same reasoning as our discussion in App.~\ref{app:A}.
Since $\Lambda_d$ is approximately the scale at which $\alpha_d$ diverges, the size of $\tilde{\Lambda}_d$ can be estimated by computing the change in this scale between the two minima.
For $SU(N_c^d)$, with $\tilde{d}$ and $\tilde{u}$ both in the fundamental representation, we find that $\tilde{\Lambda}_d \lesssim 10^{1.5} \Lambda_d$, approaching $\Lambda_d$ as $N_c^{d}\rightarrow \infty$.
This UV completion delivers the potential in Eq.~\eqref{eq:B1}.

In our warm inflation scenario, the sector containing the waterfall field $\sigma$, the dark $SU(N_c^d)$ gauge bosons, and dark quarks, is only very weakly coupled to QCD (the scattering rate between the axion and waterfall field is suppressed by $g^4 T \ll H \sim 10^{-3} T$, and the scattering rate for axions to scatter into dark gauge bosons scales as $(T/f_d)^2 T \ll H$). Thus, the waterfall field and quarks do not become thermalized, and mesons form as long as $ H \ll \Lambda_d $, $m_{\tilde{d}} \ll \Lambda_d $, which allows for $\Lambda_d < T$. 

Plugging in sample values, we find that this UV completion is natural for the majority of our parameter space. For example for $T = 10^5 \, \text{GeV}$, $Q = 50$, $g \approx 10^{-6}$, $M = 10^5 \, \text{GeV}$, $y = y' = 0.6$, $m_{\tilde{d}} = 0.1 \Lambda_d$, $m_{\tilde{u}} = 10 \, \Lambda_{d}$, $\Lambda_d \leq  10^5 \, \text{GeV}$, $f_d = 10^{10} \, \text{GeV}$, we find $\delta m_{\phi} < 0.5 \,  \text{GeV} \lesssim m_{\phi}$, $\delta M < 10^9 \, \text{GeV} < M $, and after the waterfall field acquires a vacuum expectation value $m^{\text{eff}}_{\tilde{d}}\sim 10^{17}\, \text{GeV} \gg \Lambda_d$, such that $g^2 = 0 $ today. As $T$ increases the consistency conditions become more easily satisfied.  

It is tempting to label the above $\delta m_\phi^2$ as our heavy QCD axion mass contribution $m_\phi^2$, instead of requiring that it is smaller.
However, achieving this would require a more complicated UV completion that also solves the strong CP problem, since there is no reason why the $\theta$ angle of this $SU(N_c^d)$ sector would be the same as QCD. 
If such a scenario is achievable, the resulting constraint plot (Fig.~\ref{fig:Constraints}) would have the allowed region shifted right, since the confinement scale increases after the phase transition.

\bibliographystyle{JHEP}
\bibliography{bib}

\end{document}